\documentclass[12pt,draftcls,onecolumn]{IEEEtran}
\usepackage{amsfonts}
\usepackage{amssymb}
\usepackage{mathrsfs}
\usepackage{verbatim}
\usepackage{subfigure}
\usepackage[centertags]{amsmath}
\usepackage{graphicx}  
\newtheorem{theorem}{\bf Theorem}
\usepackage{epsfig}
\usepackage{psfig}
\usepackage{subfigure}

\newtheorem{example}{Example}

\providecommand{\norm}[1]{\lVert#1\rVert}

\newtheorem{lemma}{{ Lemma}}

\title{A Class of Maximal-Rate, Low-PAPR, Non-square Complex Orthogonal Designs}
\author{Smarajit Das,~\IEEEmembership{Student Member,~IEEE} and
        B. Sundar Rajan,~\IEEEmembership{Senior Member,~IEEE}%
\thanks{This work was supported through grants to B.S.~Rajan; partly by the IISc-DRDO program on Advanced Research in Mathematical Engineering, and partly by the Council of Scientific \& Industrial Research (CSIR, India) Research Grant (22(0365)/04/EMR-II). 
Smarajit Das and B. Sundar Rajan are with the Department of Electrical Communication Engineering, Indian Institute of Science, Bangalore-560012, India. Email:\{smarajit,bsrajan\}@ece.iisc.ernet.in.}}

\begin{document}
\maketitle
\begin{abstract}
Space-time block codes (STBCs) from non-square complex orthogonal designs are bandwidth efficient when compared with those from square real/complex orthogonal designs.
Though there exists rate-$1$ ROD for any number of transmit antennas, rate-$1$ complex orthogonal designs (COD) does not exist for more than $2$ transmit antennas.
Liang (IEEE Trans. Inform. Theory, 2003) and Lu et al (IEEE Trans. Inform. Theory, 2005) have constructed a class of maximal rate non-square CODs where the rate is $\frac{1}{2}+\frac{1}{n}$ if number of transmit antennas $n$ is even and  $\frac{1}{2}+\frac{1}{n+1}$ if $n$ is odd. In this paper, we present a simple construction for maximal rate non-square CODs obtained from square CODs which resembles the construction of rate-1 non-square RODs from square RODs. These designs are shown to be amenable for  construction of a class of  generalized CODs (called Coordinate-Interleaved Scaled CODs) with low peak-to-average power ratio (PAPR) having the same parameters as the maximal rate codes. Simulation results indicate that these codes perform better than the existing maximal rate codes under peak power constraint while performing the same under average power constraint.

\end{abstract}
\begin{keywords}
MIMO, orthogonal designs, PAPR, space-time codes, transmit diversity.
\end{keywords}

\section{Introduction and Preliminaries}
There are several definitions of Orthogonal Designs (ODs) in the literature \cite{TJC,Lia,TWMS} the well known being as given in \cite{Lia}:
A {\textit{linear-processing complex orthogonal design}} (LCOD) is a $p\times n$ matrix $G$ in $k$ complex variables $x_0,x_1,\cdots,x_{k-1}$ such that each non-zero entry of the matrix is a complex linear combinations of the complex variables $x_0,x_1,\cdots,x_{k-1}$ and their conjugates $x_0^*,x_1^*,\cdots,x_{k-1}^*$ satisfying $G^\mathcal{H}G=({\vert x_0\vert}^2 +{\vert x_1\vert}^2+\cdots+{\vert x_{k-1}\vert}^2)I_n$, where $G^\mathcal{H}$ is the complex conjugate transpose of $G$ and $I_n$ is the $n\times n$ identity matrix.
An LCOD $G$ is called {\textit{complex orthogonal design}} (COD) if the non-zero entries of $G$ are the complex variables $\pm x_0,\pm x_1,\cdots,\pm x_{k-1}$ or its complex conjugates. 

To construct non-square CODs with low PAPR, we identify a subclass of LCODs which includes CODs as a special case. This is done  using the notion of coordinate interleaved complex variables \cite{KhR} which has been extensively used to construct single-symbol decodable STBCs that are not CODs.  Given two complex variables $s_1$ and $s_2$ where $s_i=s_{iI}+js_{iQ},i=1,2$, the coordinate interleaved variables corresponding to the variables $s_1$ and $s_2$, are $\hat{s}_1=s_{1I}+js_{2Q}$ and $\hat{s}_2=s_{2I}+js_{1Q}$. An LCOD is called {\textit {coordinate interleaved scaled complex orthogonal designs}} (CIS-COD) if any non-zero entry of the matrix is a variable or a coordinate interleaved variable, or their complex conjugates, or multiple of these by $\pm 1$ or $\pm \frac{1}{\sqrt{2}}$. we call an orthogonal design with the parameters $p,n$ and $k$ as stated above a $[p,n,k]$ orthogonal design or an orthogonal design of size $[p,n,k]$. The rate of a $[p,n,k]$ orthogonal design is defined to be  $\frac{k}{p}$.

Space-time block codes (STBCs) from complex orthogonal designs (CODs) have been extensively studied for square designs, since they correspond to minimum decoding delay codes.
The rate of the square CODs falls exponentially with increase in the number of transmit antennas. Specifically,
\begin{theorem}[~\cite{TiH},~\cite{ALP}]
The maximal rate of a square complex orthogonal design is given by $\frac{a+1}{n}$ where $a$ is the exponent of $2$ in the prime factorization of $n$ .
\end{theorem}
Several authors have constructed square CODs achieving maximal rate \cite{TiH, ALP}. In \cite{TiH}, the following induction method is used to construct square CODs for $2^a$ antennas, $a=2,3,\cdots$, starting from 
{\small
\begin{equation}
\label{itcod}
G_1= \left[\begin{array}{rr}
x_0   &-x_1^*      \\
x_1   & x_0^*
\end{array}\right],~
G_a=  \left[\begin{array}{rr}
G_{a-1}   & -x_{a}^*I_{2^{a-1}}    \\
x_{a}I_{2^{a-1}}   & G_{a-1}^\mathcal{H}
\end{array}\right],
\end{equation}
}
\noindent
where $G_a$ is a $2^a\times 2^a$ complex matrix. Note that $G_a$ is a square COD in $(a+1)$ complex variables $x_0,x_1,x_2,\cdots, x_{a}$.

It is clear from the above theorem that the square OD, real/complex are not bandwidth efficient and naturally one is led to study non-square orthogonal designs in order to obtain codes with high rate. It is known that \cite{TJC} there always exists a rate-1 real orthogonal design (ROD) for any number of transmit antennas and these codes are constructed from square CODs \cite{TJC}.
On the other hand, it is not known, in general, the maximal rate of complex orthogonal design which admits as entries the arbitrary linear combination of complex variables. However, it is shown by Liang \cite{Lia} that the maximal rate of  a COD, when the non-zero entries of the designs are only the variables or their conjugates with or without  negative sign, is equal to $\frac{a+1}{2a}$ when number of transmit antennas is $2a-1$ or $2a$. He has also given an explicit construction of CODs achieving this rate for any number of antennas. There is also another construction of these codes given by Lu et al \cite{LFX}. 

{\it Contributions of this paper:} The contributions of this paper may be summarized as follows: 
\begin{itemize} 
\item We present a simple construction for maximal rate non-square CODs for any large number of antennas having the same delay as that of \cite{Lia} for the number of antennas not multiples of 4 and of the same delay as that of \cite{LFX} for number of antennas multiple of 4. The construction of these CODs starts from square CODs and is very similar to the construction of rate-1 non-square RODs from square RODs of \cite{TJC}. The constructed codes are amenable for modification to codes with low PAPR.

\item Starting from the  maximal rate  codes mentioned above, we have also constructed a class of maximal rate CIS-CODs which have the same delay as that of the codes given by Liang \cite{Lia} and Lu et al \cite{LFX}, but having smaller number of zero entries, leading to codes having low peak-to-average power ratio (PAPR). These codes perform better than the known codes under peak power constraint while perform same under average power constraint. Simulation results are presented which justify this claim.
\end{itemize}

The remaining part of the paper is organized as follows: In Section \ref{sec2}, we give construction of maximal rate achieving CODs from square CODs. In Section \ref{sec3}, we give code construction CIS-CODs which has low PAPR. In Section \ref{sec4}, we provide some simulation results. Section \ref{sec5} concludes the paper. 

\section{A simple construction of maximal rate CODs from square CODs}
\label{sec2}

It is known that the maximal rate of a real orthogonal design is one for any number of transmit antennas and these codes can be constructed from square RODs. This method does not apply to the construction of maximal rate achieving non-square CODs as some of the variables in the matrix are complex conjugated.
In this section, it is shown that one can still construct maximal rate non-square CODs from square CODs if the method used for the construction of rate-$1$ RODs from square RODs is suitably modified. For this purpose, we introduce some notations:

Let $\mathbb{F}_2$ be the finite field with two elements denoted by $0$ and $1$ with addition denoted by $b_1\oplus b_2$ and multiplication denoted by $b_1b_2$ where $ b_1,b_2 \in \mathbb{F}_2.$ Let  $b_1b_2$, $b_1 +b_2$ and $\bar{b_1}$ represent respectively the logical operations of conjunction (AND), disjunction (OR) and complement or negation.  All other Boolean operations are obtained from these basic operations. For example, the exclusive-or (XOR) of $b_1,b_2$, denoted by $b_1\oplus b_2$, is equal to $\bar{b_1}b_2+\bar{b_2}b_1$. Note that
$b_1+b_2=b_1\oplus b_2 \oplus b_1b_2,~~~~ \bar{b_1}=1\oplus b_1.$ 

Let $B$ be a finite subset of the set of natural numbers with  $b$ being its largest element and $a\in\mathbb{N}$ be such that $2^a>b$. We can always identify each element of $B$ with an element of $\mathbb{F}_2^a$ using the following correspondence:
$b\in B \leftrightarrow (b_{a-1},\cdots,b_0)\in\mathbb{F}_2^a$ such that $b=\sum_{j=0}^{a-1} b_j2^j,b_j\in \mathbb{F}_2$.
The all zero vector and all one vector in $\mathbb{F}_2^a$ are denoted by $\mathbf{0}$ and $\mathbf{1}$ respectively. For $x\in B$, $\overline{x}$ and $\norm{x}$ 
represent respectively the $2's$ complement of $x$ in $\mathbb{F}_2^a$ and Hamming weight of $x$.
Let $x=(x_{a-1},\cdots,x_0),y=(y_{a-1},\cdots,y_0)$. Let 
$x\oplus y, x\cdot y$ denote the component-wise modulo-2 addition and component-wise  multiplication (AND operation)  of $x$ and $y$ respectively i.e.,
\begin{eqnarray*}
x\oplus y=(x_{a-1}\oplus y_{a-1},\cdots,x_0\oplus y_0),~~~~x\cdot y=(x_{a-1}y_{a-1},\cdots,x_0y_0).
\end{eqnarray*}
Let $Z_l=\{0,1,\cdots,l-1\}.$ For a set  $K\subset Z_{2^a}$, define
$m \oplus K:=\{m\oplus a~\vert~ a\in K\}$ for some $m\in Z_{2^a}$ and  $\vert K\vert$ be the number of elements in the set $K$. Let $A,B$ be two sets and $B\subset A$. Denote by $A\setminus B$, the set of those elements of $A$, which are not in $B$. For two integers $i,j$, we use the notation $i\equiv j$, to indicate that the difference of $i$ and $j$ is an even number. 

Let
\begin{eqnarray}
\label{tk}
\begin{array}{ccl}
R_a&=&\Big\{i\in Z_{2^a}~\Big\vert~ \norm{i}\in\{\lceil \frac{a}{2}\rceil-2,\lceil \frac{a}{2}\rceil-1,\lceil \frac{a}{2}\rceil,\lceil \frac{a}{2}\rceil+1\}\Big\} \\
C_a&=&\Big\{i\in Z_{2^a}~\Big\vert~ \norm{i}\in\{\lceil \frac{a}{2}\rceil-1,  \lceil \frac{a}{2}\rceil\}\Big\},
\end{array}
\end{eqnarray}
where $\lceil x \rceil$ is the smallest integer greater than or equal to $x.$ These two subsets of $Z_{2^a}$ play an important role subsequently in the construction of our codes. Notice that $C_a \subset R_a$ and  $C_a$ and $R_a$ possess the following  nice properties.
\begin{lemma}
\label{alp}
Let the integers $s$ and $t,$  $s \neq t,$ be such that $C_a\cap (2^s\oplus C_a)\neq\phi$ and $C_a\cap (2^s\oplus 2^t\oplus C_a)\neq\phi$. Then
  $\norm{i}+\lceil \frac{a}{2}\rceil+i_s$ is an odd number for all $i\in C_a\cap (2^s\oplus C_a)$ and $i_s+i_t=1$ for all $i\in C_a\cap (2^s\oplus 2^t\oplus C_a)$.
\end{lemma}
\begin{proof}
We have 
\begin{eqnarray*} 
\begin{array}{lcl}
C_a\cap (2^s\oplus C_a)&=&\Big\{i\in C_a~\Big\vert~ (i_s=0 \text{ and } \norm{i}=\lceil\frac{a}{2}\rceil-1)
\text{ or } (i_s=1 \text{ and } \norm{i}=\lceil \frac{a}{2}\rceil)\Big\},\\
C_a\cap (2^s \oplus 2^t\oplus C_a)&=&\Big\{i\in C_a~\Big\vert~ (i_s=0 \text{ and } i_t=1)
\text{ or } (i_s=1 \text{ and } i_t=0)\Big\}.
\end{array}
\end{eqnarray*} 
Therefore, $\forall i\in C_a\cap (2^s\oplus C_a)$, $\norm{i}+\lceil \frac{a}{2}\rceil+i_s= 2\lceil\frac{a}{2}\rceil-1$ or $2\lceil\frac{a}{2}\rceil+1$ and 
$i_s+i_t=1~\forall i\in C_a\cap (2^s\oplus 2^t\oplus C_a)$.
\end{proof}
Note that $\vert C_{2a-1}\vert=2\vert C_{2a-2}\vert=2\binom{2a-1}{a}$
and $\vert R_{2a-1}\vert=2\vert R_{2a-2}\vert=\frac{2a}{a+1}\binom{2a}{a}.$ This fact will be used subsequently.
Let $M$ be a $p\times n$ matrix in $k$ complex variables $x_0,x_1, x_2,\cdots, x_{k-1}$, such that each non-zero entry of the matrix is $x_i,x_i^*,-x_i$ or $-x_i^*$ for some $i\in\{0,1,\cdots,k-1\}$. 
If $M(i,j)\neq 0$, then we write $\vert M(i,j)\vert=l$ whenever $M(i,j)\in \{\pm x_l,\pm x_l^*\}$ for some $l\in Z_k$.
It need not be true that the matrix $M$ would be a COD. For example, 
 {\small $\begin{bmatrix}
    x_0   &x_1    \\
   x_1^*  &x_0^*  
\end{bmatrix}
$} is not a COD.

Let $M_2$ be a $2\times 2$ submatrix which is constructed from $M$ by choosing two rows and two columns of $M$. The matrix $M_2$ is called {\textit{proper}} if
\begin{itemize}
\item None of the entries of $M_2$ is zero and
\item It contains exactly two distinct variables.
\end{itemize}
\begin{example}
For the  matrix in three complex variables $x_0,x_1$ and $x_2,$ given by {\footnotesize$
\left[\begin{array}{rrrrr}
    x_0   &-x_1^* &-x_2^*  & 0     \\
    x_1   & x_0^* & 0      &-x_2^*     \\
    x_2   & 0     & x_0^*  & x_1^*     \\
     0    & x_2   &-x_1    & x_0
 \end{array}\right]$ }
\noindent while the sub-matrix 
$\begin{bmatrix}
    x_0   &-x_1^*    \\
    x_1   & x_0^*   
  \end{bmatrix}$ 
is {\textit{proper}} the submatrix 
$ \begin{bmatrix}
    x_2   & 0 \\
    0     & x_2 
  \end{bmatrix}$
is not.
\end{example}
The following lemma gives a characterization of CODs in term of proper $2\times 2$ matrices whose proof is straight forward from the properties of CODs \cite{Lia}.
\begin{lemma}
\label{propercomplex}
Let $M$ be a $p\times n$ matrix in $k$ complex variables $x_0,x_1, x_2,\cdots, x_{k-1}$, such that each non-zero entry of the matrix is $\pm x_i$ or $\pm x_i^*$ for some $i\in Z_k$. Then following two statements are equivalent:\\
 1) $M$ is a COD.\\
 2) (i) Each variable appears exactly once along each column of $M$ and at most once along each row of $M$,\\
\noindent (ii) If for some $i,i^\prime,j,j^\prime$, $M(i,j)\neq 0$, $M(i,j^\prime)\neq 0$, $M(i^\prime,j^\prime)\neq 0$
         and $\vert M(i,j)\vert=\vert M(i^\prime,j^\prime)\vert$, then $M(i^\prime,j)\neq 0$ and $\vert M(i,j^\prime)\vert=\vert M(i^\prime,j)\vert$,\\
   (iii) Any proper $2\times 2$ sub-matrix of $M$ is a COD.
\end{lemma}
\subsection{A simple construction of  maximal-rate CODs} 
 \label{subsec4-1}
 In this subsection, we construct maximal-rate CODs for $t$ transmit antennas from a square COD $G_{t-1}$ which is given in \eqref{itcod}, which are amenable for extension  (in the following section) to low PAPR CODs without the rate and the delays getting changed.  
The following two lemmas Lemma \ref{lem3} and Lemma \ref{M_2} will be useful in constructing the desired CODs. In the proof of Lemma \ref{lem3}, we make use of the following two facts:
\noindent
(i) Let $N_i^{(a)}$ be the set of row indices of the non-zero entries of the $i$-th column of $G_a$.
It is known \cite{DaR1} that
\begin{equation}
\label{eni}
N_i^{(a)}=\{i\}\cup\{i \oplus 2^j~\vert~ j=0 \mbox { to } a-1\}.
\end{equation}
\noindent
(ii) For $t$ number of transmit antennas, let $k_t$ and $p_t$ be the number of complex variables and the decoding delay of the maximal-rate non-square CODs constructed by \cite{Lia}. Then, for the later use, we have
\begin{eqnarray}
\label{vtut}
\begin{array}{rlrll}
k_t&=\binom{2l-1}{l}, & p_t&=\frac{l}{l+1}\binom{2l}{l}, &\mbox{ if } t=2l-1, \\
k_t&=2k_{t-1},        & p_{t}&=2p_{t-1},                  &\mbox{ if } t=2l.
\end{array}
\end{eqnarray}
\begin{lemma}
\label{lem3}
For an integer $a,$ let  $p_{a+1},k_{a+1}$ be as defined in \eqref{vtut}. Then, 
a COD of size $[p_{a+1},k_{a+1},a+1]$ is constructable.
\end{lemma}
\begin{proof}
We give an explicit construction of a COD of size $[p_{a+1},k_{a+1},a+1]$ for all $a$.
This matrix is constructed from $G_a$ by removing some of its columns. Recall that the rows and columns of $G_a$ are indexed by the elements of the set $Z_{2^a}$. Let $M$ be a matrix formed by the columns of $G_a$ which are in $C_a$ i.e., $M$ contains the $i$-th column of $G_a$ if $i\in C_a$. The number of rows in the matrix $M$ is $2^a$. Now we determine those rows of $M$ which contain at least one non-zero entry. These rows are those rows of $G_a$ whose indices lie in the set $\cup_{i\in C_a}{N_i}^{(a)}$ where ${N_i}^{(a)}$ is given by \eqref{eni}.
Now 
\begin{eqnarray*}
\begin{array}{ccl}
\cup_{i\in C_a}{N_i}^{(a)}&=&\Big\{i\in Z_{2^a}~\Big\vert~ \norm{i}\in\{\lceil \frac{a}{2}\rceil-2,\lceil \frac{a}{2}\rceil-1,\lceil \frac{a}{2}\rceil,\lceil \frac{a}{2}\rceil+1\}\Big\}.
\end{array}
\end{eqnarray*}
Therefore, $\cup_{i\in C_a}{N_i}^{(a)}=R_a$.
Removing those rows of $M$ which contain only zeros, we get a matrix which has $\vert C_a\vert =k_{a+1}$ columns and $\vert R_a\vert=p_{a+1}$ rows.
\end{proof}
Let the elements of $C_a$ be $c_i, i=0 \text{ to }k_{a+1}-1$ such that $c_0<c_1<\cdots<c_{k_{a+1}-1}$ and that of $R_a$ be  $r_i, i=0 \text{ to }p_{a+1}-1$ satisfying $r_0<r_1<\cdots<r_{p_{a+1}-1}$. Define $f:Z_{p_{a+1}}\mapsto R_a$ and $g:Z_{k_{a+1}}\mapsto C_a$  as follows:
\begin{eqnarray}
\label{functionfg}
\begin{array}{r}
  f  : Z_{p_{a+1}}\rightarrow R_a\\
       i     \mapsto       r_i 
\end{array},
~~~~~~
\begin{array}{r}
  g  : Z_{k_{a+1}}\rightarrow C_a\\
       i     \mapsto   c_i 
\end{array}.
 \end{eqnarray}
Note that both $f$ and $g$ are bijective maps.  Also, we use the following notation:
\begin{eqnarray*}
2_+^x&=&
    \begin{cases}
       2^x    & \mbox{ if } x \geq 0,\\
       0      & \mbox{ if } x=-1. 
    \end{cases}
\end{eqnarray*}
(The situation where $x < -1$  does not arise.)

For any matrix of size $n_1\times n_2,$ the rows and columns of the matrix are indexed by the elements of $Z_{n_1}$ and $Z_{n_2}$ respectively. For a $[p,n,k]$ COD $Q$ in $k$ complex variables $x_0,x_1,\cdots,x_{k-1},$  we define three maps $\lambda_Q:Z_p\times Z_n\rightarrow Z_k$ and ${\mu}_Q,{\tau}_Q:Z_p\times Z_n\rightarrow \{1,-1\}$ associated to $Q$ as follows:\\
If $Q(i,j)\neq 0$, then
\begin{enumerate}
\item $\lambda_Q(i,j)=c$ if $Q(i,j)\in\{\pm x_c, \pm x_c^*\}$ for some $c\in Z_k$,
\item $\mu_Q(i,j)=1$ if $Q(i,j)=x_c$ or $x_c^*$ and $-1$ if $Q(i,j)=-x_c$ or $-x_c^*$,
\item $\tau_Q(i,j)=1$ if $Q(i,j)=x_c$ or $-x_c$ and $-1$ if $Q(i,j)=x_c^*$ or $-x_c^*$. \end{enumerate}
Sometimes, for notational simplicity,  we write $\lambda,\tau$ and $\mu$ for $\lambda_Q,\tau_Q$ and $\mu_Q$ respectively when the underlying $Q$ is clear from the context. Let $P_a=\{0,1,2^1\cdots,2^{a-1}\}$. If $Q$ is a square COD $G_a$ defined in \eqref{itcod},
then $G_a(i,j)\neq 0$ if and only if $(i\oplus j)\in P_a$.
When $G_a(i,j)\neq 0$, we have
\begin{eqnarray*}
\lambda_{G_a}(i,j)&=&l \mbox{ if } i\oplus j=2_+^{l-1},\\
\tau_{G_a}(i,j)&=&
    \begin{cases}
       (-1)^{\norm{j}}  \hspace{16pt}        & \mbox{ if } i\oplus j=0,\\
        (-1)^{j_{l-1}}          & \mbox{ if }i\oplus j=2^{l-1},l\neq 0,\\
    \end{cases}, \\
\mu_{G_a}(i,j)&=&
    \begin{cases}
       1          & \mbox{ if } i\oplus j=0,\\
     (-1)^{\norm{j\cdot \overline{2^{l-1}}}} & \mbox{ if }i\oplus j=2^{l-1}, l\neq 0.\\
    \end{cases}
     \end{eqnarray*}

We denote the COD of size $[p_{a+1},k_{a+1},a+1]$ constructed in the proof of Lemma \ref{lem3} above by $\widetilde{H}_a$. Note that $\widetilde{H}_a(i,j)\neq 0$ if $f(i)\oplus g(j)\in P_a$.
On the matrix $\widetilde{H}_a$, the following operation is performed.
When $a$ is $1$ or $2$ (modulo $4$),
we substitute $x_0$ in $\widetilde{H}_a$ by $x_0^*$ and $x_0^*$ by $x_0$. Then, we have
\begin{eqnarray}
\label{tildeHaadj}
\tau_{\widetilde{H}_a}(i,j)&=&
    \begin{cases}
       (-1)^{\norm{g(j)}+\lceil \frac{a}{2}\rceil}          & \mbox{ if } f(i)\oplus g(j)=0\\
        (-1)^{{g(j)}_{l-1}}          & \mbox{ if }f(i)\oplus g(j)=2^{l-1},l > 0
    \end{cases}.
     \end{eqnarray}
The matrix $\widetilde{H}_a$ has a very nice property, namely all the complex variables that lie in the same row are either complex-conjugated or none of them is complex-conjugated. We call a COD with this property a {\textit{conjugation-separated}} COD. As an example, observe that the COD $\widetilde{H}_4$ given on the left side of \eqref{h4h4} is conjugation-separated.

\begin{lemma}
\label{M_2}
$\widetilde{H}_a$ is conjugation-separated.
\end{lemma}
\begin{proof}
Let $i\in Z_{p_{a+1}},j,j^\prime \in Z_{k_{a+1}},j\neq j^\prime$ such that $\widetilde{H}_a(i,j)\neq 0$ and $\widetilde{H}_a(i,j^\prime)\neq 0$.
 We show that $\tau(i,j)=\tau(i,j^\prime)$ i.e., $\tau(i,j)\tau(i,j^\prime)=1$ where we write $\tau$ for $\tau_{\widetilde{H}_a}$.\\
Now $f(i)\oplus g(j),f(i)\oplus g(j^\prime)\in P_a$ as $(i,j)$-th and $(i,j^\prime)$-th entry of $\widetilde{H}_a$ is non-zero and $f(i)\oplus g(j)\neq f(i)\oplus g(j^\prime)$.\\
We consider two cases namely (i) $f(i)\oplus g(j)=0$ and (ii) $f(i)\oplus g(j)\neq0$.
In both the cases, $f(i)\oplus g(j^\prime) \neq 0$. Let $f(i)\oplus g(j^\prime)=2^l$ for some $l\in Z_a$.\\
{\bf Case (i)  $f(i)\oplus g(j)=0$:}
In this case, $g(j)=f(i)$, hence $g(j)=g(j^\prime)+2^l$. We have $g(j)\in C_a\cap 2^l\oplus C_a$.
Now $\tau(i,j)=(-1)^{\norm{g(j)}+\lceil \frac{a}{2}\rceil},\tau(i,j^\prime)=(-1)^{{g(j^\prime)}_{l}}$.
It is enough to show that $\norm{g(j)}+\lceil \frac{a}{2}\rceil+{g(j^\prime)}_{l}$ is an even number.
But $\norm{g(j)}+\lceil \frac{a}{2}\rceil+{g(j)}_{l}$ is an odd number by Lemma~\ref{alp} for all $g(j)\in C_a\cap 2^l\oplus C_a$ and ${g(j^\prime)}_{l}={g(j)}_{l}+1$ if ${g(j)}_{l}=0$ and $ {g(j^\prime)}_{l}={g(j)}_{l}-1$ if ${g(j)}_{l}=1$.

{\bf Case (ii)  $f(i)\oplus g(j)\neq 0$:}
Let $f(i)\oplus g(j)=2^m$ for some $m\in Z_a$. Note that $m\neq l$ as $j\neq j^\prime$.
We have $g(j)=g(j^\prime)\oplus 2^l\oplus 2^m$ i.e., $g(j)\in C_a\cap (2^l\oplus 2^m\oplus C_a)$.
Now $\tau(i,j)=(-1)^{{g(j)}_{m}},\tau(i,j^\prime)=(-1)^{{g(j^\prime)}_{l}}$.
Now ${g(j)}_m +{g(j)}_l=1$ for all ${g(j)}\in C_a\cap 2^l\oplus 2^m\oplus C_a$.
Therefore, ${g(j)}_m +{g(j^\prime)}_l$ is an even number and hence $\tau(i,j)\tau(i,j^\prime)=1$.
\end{proof}

\noindent Now, we obtain the maximum rate achieving COD of size $[p_{a+1},a+1,k_{a+1}]$, denoted by $H_a$ as follows. Let the complex variables in the matrix $H_a$ be $y_0,y_1,\cdots,y_{k_{a+1}-1}$.
$H_a(i,j)=0$ if $x_j$ or its variant ($-x_j, x_j^*,-x_j^*$) is absent in the $i$-th row of $\widetilde{H}_a$, otherwise

{\footnotesize
\begin{eqnarray}
\label{maxiamlH}
 H_a(i,j)=
    \begin{cases}
\begin{array}{rll}
 y_l    & \text{ if } &\widetilde{H}_a(i,l)= x_j, \\
-y_l    & \text{ if }&\widetilde{H}_a(i,l)=-x_j, \\
 y_l^*  & \text{ if }&\widetilde{H}_a(i,l)= x_j^*,\\
-y_l^*  & \text{ if }&\widetilde{H}_a(i,l)=-x_j^*.
\end{array}
\end{cases}
\end{eqnarray}
}
\begin{theorem}
$H_a$ is a COD of size $[p_{a+1},a+1,k_{a+1}]$ for all positive integers $a$.
\end{theorem}
\begin{proof}
We use Lemma~\ref{propercomplex} to prove the theorem. By the construction of the matrix $H_a,$ each column of matrix $H_a$ contains all the variables namely $y_0,y_1,\cdots,y_{k_{a+1}-1}$ exactly once and each variable appears at most once in any row of $H_a$. Next, assuming that $H_a(i,j),H_a(i^\prime,j^\prime),H^\prime_a(i,j^\prime)$ are non-zero and $\vert H_a(i,j)\vert=\vert H_a(i^\prime,j^\prime)\vert$, we show that  $H_a(i^\prime,j)$ is non-zero and $\vert H_a(i^\prime,j)\vert=\vert H_a(i,j^\prime)\vert$.
Let $\vert H_a(i,j)\vert=\vert H_a(i^\prime,j^\prime)\vert=l$.
Then, $\vert\widetilde{H}_a(i,l)\vert=j$ and $\vert\widetilde{H}_a(i^\prime,l)\vert=j^\prime$. 
Let $\vert H_a(i,j^\prime)\vert=l^\prime$ as $H_a(i,j^\prime)\neq 0$.
We have $\vert\widetilde{H}_a(i,l^\prime)\vert=j^\prime$. 
Therefore, $\vert\widetilde{H}_a(i,l^\prime)\vert=\vert\widetilde{H}_a(i^\prime,l)\vert$.
Hence $\widetilde{H}_a(i^\prime,l^\prime)\neq 0$ and  $\vert\widetilde{H}_a(i,l)\vert=\vert\widetilde{H}_a(i^\prime,l^\prime)\vert=j$.
We have $H_a(i^\prime,j)\neq 0$ and $\vert H_a(i^\prime,j)\vert=l^\prime$.
But $\vert H_a(i,j^\prime)\vert=l^\prime$. Therefore, $\vert H_a(i^\prime,j)\vert=\vert H_a(i,j^\prime)\vert$.\\
It remains to prove that any proper $2\times 2$ sub-matrix $M_H$ of $H_a$ is a COD.
Let $r,s,t$ and $u$ be binary variables which take value either $0$ or $1$ and $y$ be a complex variable. Let $y^{(l)}=y$ if $l=0$ and $y^*$ if $l=1$.
Using this, the matrix $M_H$ formed by two rows $r_1, r_2$, with $r_2> r_1$ and two columns $c_1,c_2$, $c_2> c_1$ of $H_a$ 
and the corresponding matrix  $M_{\widetilde{H}}$ of $\widetilde{H}_a$ are given
by  
\begin{equation}
  M_H=\left[\begin{array}{cc}
    ay_i^r   & by_j^s      \\
    cy_j^t   & dy_i^u
    \end{array}\right], ~~~ 
M_{\widetilde{H}}=\left[\begin{array}{cc}
    ax_{c_1}^r   & bx_{c_2}^s     \\
    dx_{c_2}^u   & cx_{c_1}^t
    \end{array}\right]. 
\end{equation}
where $a,b,c,d \in \{\pm 1 \}.$
But $M_{\widetilde{H}}$ is a COD of size $[2,2,2]$. So $abcd=-1$.  By Lemma ~\ref{M_2}, we have $r=s=0$ and $u=t=1$ or $r=s=1$ and $u=t=0$. 
Hence, $M_H$ is a COD.
\end{proof}
\begin{example}
CODs $\widetilde{H}_4$ of size $[15,10,5]$ and $H_4$ of size $[15,5,10]$ are shown in Fig. \ref{fig3}. 
\end{example}
Note that the construction of the maximum rate COD $H_a$  of size $[p_{a+1},a+1,k_{a+1}]$ involves two steps:\\
(i) first, a non-square COD $\widetilde{H}_a$ of size $[p_{a+1},k_{a+1},a+1]$ is constructed from $G_a$ and then \\
(ii) $H_a$ is obtained from $\widetilde{H}_a$.

In the following subsection, we present another construction for the same CODs $H_a.$
\subsection{A direct construction of $H_a$} 
We now give a direct construction of maximum rate CODs for any number of transmit antennas. We define a $p_{a+1}\times {(a+1)}$ matrix $H^\prime_a$ in $k_{a+1}$ complex variables $y_0,y_1,\cdots,y_{k_{a+1}-1}$ as follows:\\
Let $H^\prime_a(i,j)$ be the $(i,j)$-th element of $H^\prime_a$.
We now define $\lambda_{H^\prime_a},\mu_{H^\prime_a}$ and $\tau_{H^\prime_a}$ for the matrix $H^\prime_a$:  
Let $H^\prime_a(i,j)=0$ if $f(i)\notin 2_+^{j-1}\oplus C_a$. For $f(i)\in 2_+^{j-1}\oplus C_a$, define
\begin{eqnarray}
\label{Hprimea}
\begin{array}{lcl}
\lambda_{H^\prime_a}(i,j)&=&g^{-1}(f(i)\oplus 2_+^{j-1}) \\
\tau_{H^\prime_a}(i,j)&=&
    \begin{cases}
       (-1)^{\norm{f(i)}+\lceil\frac{a}{2}\rceil}          & \mbox{ if } j=0,\\
        (-1)^{1+ \norm{f(i)\cdot 2^{j-1}}}          & \mbox{ if }j\in\{1,2,\cdots,a\}\\
    \end{cases}\\
\mu_{H^\prime_a}(i,j)&=&
    \begin{cases}
       1          & \mbox{ if } j=0,\\
      (-1)^{1+ \norm{ f(i)\cdot \overline{2^{j-1}}}} & \mbox{ if }j\in\{1,2,\cdots,a\}\\
    \end{cases}
\end{array}
     \end{eqnarray}
for $0\leq i\leq p_{a+1}-1$ and $0\leq j\leq a$.
\begin{theorem}
\label{thm3}
$H^\prime_a$ is a non-square COD of size $[p_{a+1},k_{a+1},a+1]$ where 
$$k_{a+1}=\binom{2l-1}{l},~~p_{a+1}=\frac{l}{l+1}\binom{2l}{l}~~~~ \mbox{ if } a=2l-2~~~\mbox{  and  } $$
$$k_{a+1}=2k_{a}, ~~p_{a+1}=2p_{a} ~~~~ \mbox{ if } a=2l-1.$$
\end{theorem}
\begin{proof}
The proof is given in Appendix \ref{append1}.
\end{proof}
\subsection{Construction of maximal rate codes  with reduced delay for number of transmit antennas a multiple of 4}
\label{subsec4-2}
In the previous two subsections, we have only concentrated on the construction of maximal rate achieving codes which need not be delay optimal. 
It has been shown by Lu et al \cite{LFX} that whenever number of transmit antennas is a multiple of $4$, one can reduce the delay of the maximal rate achieving codes given in Liang et al \cite{Lia} by $50\%$. 
We now provide a simple construction of the codes for multiple of four antennas with the decoding delay as above.
Let this non-square COD for $4m$ transmit antennas be denoted by $\hat{H}_{4m}$.
The design $H^\prime_{4m-2}$ constitutes all the columns of $\hat{H}_{4m}$ except the last column.
 Note that the number of rows and complex variables in both $\hat{H}_{4m}$
and $H^\prime_{4m-2}$ are same whereas the number of columns in $\hat{H}_{4m}$ is one more than that of $H^\prime_{4m-2}$. Let $\mathbf{1}$ denote the all $1$ vector in the vector space $\mathbb{F}_2^{4m-2}$ (over $\mathbb{F}_2$) and 
$\hat{\mathbf{1}}=2^0+2^2\cdots +2^{4m-4}$.\\
We now define $\lambda_{\hat{H}_{4m}},\mu_{\hat{H}_{4m}}$ and $\tau_{\hat{H}_{4m}}$
for $\hat{H}_{4m}$ as follows:
\[
\chi_{\hat{H}_{4m}}(i,j)=\chi_{H^\prime_{4m-2}}(i,j),\chi=\lambda, \mu, \tau  \mbox { for } 0\leq i\leq p_{4m-1}-1 \mbox { and } 0\leq j\leq 4m-2. 
\]
We construct the last column of $\hat{H}_{4m}$ as follows. Let $\hat{H}_{4m}(i,4m-1)=0$ if $f(i)\notin \mathbf{1}\oplus C_{4m-2}$.
When $H^\prime_a(i,j)\neq 0$, define
   \begin{eqnarray}
\label{Hhat4m1}
\begin{array}{lcl}
\lambda_{\hat{H}_{4m}}(i,4m-1)&=& g^{-1}(f(i)\oplus \mathbf{1}),\\
\mu_{\hat{H}_{4m}}(i,4m-1)&=&(-1)^{1+ \norm{f(i)\cdot \hat{\mathbf{1}}}},\\
\tau_{\hat{H}_{4m}}(i,4m-1)&=&-1.
\end{array}
\end{eqnarray}
\begin{theorem}
\label{thm4}
$\hat{H}_{4m}$ is a COD of size $[\frac{2m}{2m+1}\binom{4m}{2m}, 4m, \binom{4m-1}{2m}   ].$
\end{theorem}
\begin{proof}
The proof is given in Appendix \ref{append2}.
\end{proof}
Thus, the decoding delay and the rate of the non-square CODs for $(4m-1)$ and $4m$ transmit antennas 
given by $H^\prime_{4m-2}$ and $\hat{H}_{4m}$ respectively,
are identical. 
As an example, the rate-5/8 non-square COD $\hat{H}_8$ of size $[56,8,35]$ is given in Fig. \ref{c1fig3}.
\section{A class of CIS-CODs  with low PAPR} 
\label{sec3}
Besides the rate, diversity and decodability of space-time codes, low PAPR of a code is an important parameter. It is desirable to construct code with low PAPR for ease of practical implementation of these codes in wireless communication system. One possible way to construct a code with low PAPR is to reduce the number of zeros in an existing code without increasing the signaling complexity \cite{DaR2} significantly. This has been discussed elaborately for square CODs in \cite{DaR2}. In this section, we obtain a class of maximal-rate non-square CIS-CODs with low PAPR, the techniques used for which are completely different from those employed for square CODs and non-trivial.
 
As the maximal rate of a COD for $2t-1$ or $2t$ transmit antennas is 
$\frac{t+1}{2t}$, the fraction of zeros in the codeword matrix is given by $1-\frac{t+1}{2t}=\frac{t-1}{2t}$.
For example, consider the two codes (i) and (ii) given by \eqref{maximal3} for three transmit antennas 

{\small
\begin{eqnarray}
\label{maximal3}
(i)\left[\begin{array}{rrrr}
    y_0   &-y_1^*  &-y_2^*\\
    y_1   & y_0^*  & 0    \\
    y_2   & 0      & y_0^*\\
     0    & y_2    &-y_1   
\end{array}\right],~
(ii)\left[\begin{array}{rrrr}
    x_0   &-x_1^*  &-\frac{x_2^*}{\sqrt{2}}\\
    x_1   & x_0^*  &-\frac{x_2^*}{\sqrt{2}}\\
    \frac{x_2}{\sqrt{2}}   & \frac{x_2}{\sqrt{2}}    & x_{1I}-jx_{0Q}\\
    \frac{x_2}{\sqrt{2}}   &-\frac{x_2}{\sqrt{2}}    & x_{0I}-jx_{1Q}
\end{array}\right],~
(iii)\left[\begin{array}{rrrr}
  \frac{y_0+y_1}{\sqrt{2}}   &\frac{-y_1^*+y_0^*}{\sqrt{2}}  &\frac{-y_2^*}{\sqrt{2}} \\
   \frac{y_0-y_1}{\sqrt{2}}  &\frac{-y_1^*-y_0^*}{\sqrt{2}}  &\frac{-y_2^*}{\sqrt{2}} \\
   \frac{y_2}{\sqrt{2}}      & \frac{y_2}{\sqrt{2}}          &\frac{y_0^*-y_1}{\sqrt{2}} \\
     \frac{y_2}{\sqrt{2}}    &\frac{-y_2}{\sqrt{2}}          &\frac{y_0^*+y_1}{\sqrt{2}}   
\end{array}\right].
\end{eqnarray}
}
\noindent where $x_{iI},x_{iQ}$ are the in-phase and the quadrature component of $x_i,i=0,1$ respectively.
The code (i) contains three zeros which amounts to one-forth of the total number of entries of the matrix while none of the entries in the code (ii) is zero. Moreover, the code (i) is a COD while the code (ii) is a CIS-COD which is not a COD.

Let $H^\prime_n$ be the matrix defined in \eqref{Hprimea} for $n+1$ number of antennas. The rows and columns of $H^\prime_n$ are indexed be the elements of $Z_{p_{n+1}}=\{0,1,\cdots,(p_{n+1}-1)\}$ and $Z_{n+1}=\{0,1,\cdots,n\}$ respectively. Let $Ro_i,Ro_j$ be the $i$-th and the $j$-th row vector of $H^\prime_n$ respectively and $l$ be a fixed integer between $1$ and $2^{n}-1$. 
For a chosen value of $l,$ following operations are defined on the rows of $H^\prime_n$: 
\begin{eqnarray}
\label{row_oper}
\begin{array}{l}
Ro_i\leftarrow \frac{1}{\sqrt{2}}(Ro_i+Ro_j), ~ Ro_j
\leftarrow \frac{1}{\sqrt{2}}(Ro_i-Ro_j)
\mbox { if } f(i)\oplus f(j)=l, 
\end{array}
\end{eqnarray}
where the map $f$ is given by \eqref{functionfg}.
We say that the $i$th row and the $j$-th row of $H^\prime_n$ form a $l-$pair if $f(i)\oplus f(j)=l$.
If we apply the above operations on the code (i) of \eqref{maximal3}, with $l=1$, we get the code (iii) of \eqref{maximal3}. This matrix is a LCOD which is not a COD. However, any non-zero entry of the matrix is a linear combination of at most two variable.
Note that the variable $y_2$ appears twice  in all the columns of the code (iii) and it eliminates the zeros from the code (i). The code (ii) of \eqref{maximal3} is obtained from the code (iii) by substituting 
$y_0$ with $\frac{x_0+x_1}{\sqrt{2}}$, $y_1$ with $\frac{x_0-x_1}{\sqrt{2}}$ and $y_2$ with $x_2$. 
Observe that any non-zero entry of the code (iii) consists of a single variable or two distinct variables. In the first case, we say the variable is isolated while in the latter, we say corresponding two variables  form a $l-$pair. One striking property of the above matrix is the following: if a variable, say $x_2$, appears alone in any column of the matrix, it remains so in the remaining columns of the matrix too. Similarly, if two variables, say $x_0$ and $x_1$, form a pair in a column of the matrix, then they always appear together in all the columns of the matrix. In other words, it never happens that two variables $x_0$ and $x_1$ form a pair in one column while $x_0$ and $x_2$ also form a pair in another column of the matrix. It is this property of the code (iii) that enables us to construct the code (ii).
Any entry of the latter matrix consists of a single variable or a co-ordinate interleaved variable. The reason why we prefer the latter matrix over the preceding matrix is that the latter has lesser signaling complexity than that of the former.
We will see that this property holds for the maximal rate codes given in this paper. 

Let $l\in Z_{2^n}$ and $M_n(l)$ be the matrix obtained after performing the row operations defined by \eqref{row_oper} on $H^\prime_n$. Any non-zero entry of the matrix $M_n(l)$ is of the form $\pm y_i, \pm y_i^*, \pm y_i\pm y_j, \pm y_i\pm y_j^*,\pm y_i^*\pm y_j^*, i,j\in Z_{k_{n+1}}, ~i\neq j$ scaled by $1$ or $\frac{1}{\sqrt{2}}.$ 
\begin{lemma}
\label{pair}
Let $l$ and $M_n(l)$ be as above and $i,j\in Z_{k_{n+1}}, i\neq j$. Then $y_i$ forms a $l-$pair with $y_j$ in any column of $M_n(l)$ if and only if $g(j)=g(i)\oplus l$ or $y_i$ is  isolated if and only if $g(i)\oplus l\notin C_n$,
where $g$ is the map defined in 
\eqref{functionfg}.
\end{lemma}
\begin{proof}
Suppose $y_i$ and $y_j$ form a pair in $m$-th column of $M_n(l)$. 
Without loss of generality, we can assume that $H^\prime_n(k,m)$ contains $y_i$ whereas $H^\prime_n(k^\prime,m)$ contains $y_j$ for some $k,k^\prime\in Z_{p_{n+1}}$ and $f(k)\oplus f(k^\prime)=l$.
Now, by \eqref{Hprimea}, we have $\lambda_{H^\prime_n}(k,m)=i$ and $\lambda_{H^\prime_n}(k^\prime,m)=j$.
We have $f(k)\oplus 2_+^{m-1}=g(i)$ and $f(k^\prime)\oplus 2_+^{m-1}=g(j)$.
Therefore $g(i)\oplus g(j)=l$. \\
Now assume that $g(i)\oplus g(j)=l$.
We show that $y_i$ and $y_j$ form a pair in all the columns of $M_n(l)$.
Let $m\in Z_{n+1}$.
For some $k$ and $k^\prime$, we have $\lambda_{H^\prime_n}(k,m)=i$ and $\lambda_{H^\prime_n}(k^\prime,m)=j$ i.e.,
$f(k)\oplus 2_+^{m-1}=g(i)$ and $f(k^\prime)\oplus 2_+^{m-1}=g(j)$.
We have $f(k)\oplus f(k^\prime)=l$ and hence $y_i$ and $y_j$ form a pair in the m-th column of $H^\prime_n$.\\
Similarly, one can prove the second part of the statement.
\end{proof}
Note that the number of zeros in the design matrix $M_n(l)$ is less than that of $H^\prime_n$ and depends on $l$.  Let $F_{n+1}(l)$ be the ratio of number of zeros to the total number of entries in $M_n(l)$. The following theorem gives a closed-form expression for $F_{n+1}(l)$ for an arbitrary value of $l.$
\begin{theorem}
\label{thm5}
Let $n$ be a positive integer and $l\in Z_{2^n},l\neq 0$. Then there exists an LCOD for $n+1$ transmit antennas where the rate $R_{n+1}$ and the fraction of zeros $F_{n+1}(l)$ are given by

{\footnotesize
\begin{eqnarray*}
R_{n+1}=
    \begin{cases}
\frac{1}{2}+\frac{1}{n+2}   & \text{ if $n$ is even }\\
\frac{1}{2}+\frac{1}{n+1}   & \text{ if $n$ is odd }
  \end{cases},~~
F_{n+1}(l)=
    \begin{cases}
\frac{1}{2}-\frac{1}{n+2}-\frac{a}{\frac{(n+1)(n+2)}{(n+4)}\binom{n+2}{\frac{n+2}{2}}} & \text{ if $n$ is even }\\
\frac{1}{2}-\frac{1}{n+1}-\frac{b}{\frac{2(n+1)^2}{(n+3)}\binom{n+1}{\frac{n+1}{2}}} & \text{ if $n$ is odd, $\omega_l = n$ }\\
\frac{1}{2}-\frac{1}{n+1} & \text{ if $n$ is odd, $\omega_l \neq n$}
\end{cases}\\
\end{eqnarray*}
where
\begin{eqnarray*}
a&=&\binom{w_l}{\Big\lceil\frac{w_l+1}{2}\Big\rceil}\binom{n-w_l+1}{ \frac{n}{2}-\Big\lceil\frac{ w_l}{2}\Big\rceil}
+w_l \binom{2\Big\lceil\frac{w_l}{2}\Big\rceil-1}{\Big\lceil\frac{w_l}{2}\Big\rceil}\binom{n-2\Big\lceil\frac{w_l}{2}\Big\rceil+2)}{\frac{n}{2} -\Big\lceil\frac{ w_l}{2}\Big\rceil+2}\\
 & & +(n-w_l) \binom{2\Big\lceil\frac{w_l}{2}\Big\rceil}
{\Big\lceil\frac{w_l}{2}\Big\rceil+1} \binom{n-2\Big\lceil\frac{w_l}{2}\Big\rceil+1}{ \frac{n}{2}-\Big\lceil\frac{ w_l}{2}\Big\rceil},\\
b&=&2\binom{w_l}{\Big\lceil\frac{w_l+1}{2}\Big\rceil}\binom{n-w_l}{ \frac{n+1}{2}-\Big\lceil\frac{ w_l+1}{2}\Big\rceil+1}+
w_l \binom{2\Big\lceil\frac{w_l}{2}\Big\rceil}{\Big\lceil\frac{w_l}{2}\Big\rceil}
\binom{n-2\Big\lceil\frac{w_l}{2}\Big\rceil+1)}{ \frac{n+1}{2}-\Big\lceil\frac{ w_l}{2}\Big\rceil+1}\\
 & & +(n-w_l) \binom{2\Big\lceil\frac{w_l}{2}\Big\rceil}
{\Big\lceil\frac{w_l}{2}\Big\rceil+1} \binom{n-2\Big\lceil\frac{w_l}{2}\Big\rceil+1}{\frac{n+1}{2}-\Big\lceil\frac{ w_l}{2}\Big\rceil},
\end{eqnarray*}
}
and $w_l$ is the Hamming weight of $l$.
\end{theorem}
\begin{proof}
The proof is given in Appendix \ref{append3}.
 \end{proof}
The matrix $M_n(l)$ is an LCOD where any non-zero entry contains at most two complex variables. Here any non-zero entry of $M_n(l)$ contains a variable, say $y_i,i\in Z_{k_{n+1}}$ if $g(i) \in C_n\setminus (l\oplus C_n)$, or two variables $y_i,y_j,i,j\in Z_{k_{n+1}},i\neq j$, if  $g(i),g(j)\in C_n\cap l\oplus C_n$ and $g(i)\oplus g(j)=l$. In the first case, $y_i$ is isolated and in the second case, $y_i$ forms a $l-$pair with $y_j$.
Now we construct a CIS-COD from $M_n(l)$ as follows:
If $y_i$ is isolated in the matrix $M_n(l)$, we replace it by $x_i$ and if $y_i$ and $y_j$ form a $l-$pair, then we substitute $y_i$ with $\frac{x_i+x_j}{\sqrt{2}}$ and $y_j$ with $\frac{x_i-x_j}{\sqrt{2}}$.
We denote the CIS-COD for $n$ transmit antennas constructed as above with $l=1$ by $L_n$.

For $3$ transmit antennas, we have already constructed a CIS-COD from LCOD as given by code (ii) of \eqref{maximal3}. Note that this code has no zero entry as expected.
The fraction of zeros in $L_4$ and $L_5$ are $0$ and $\frac{8}{75}$ respectively and the corresponding codes are given in Fig. \ref{c1fig4}.

From Theorem \ref{thm5}, it is clear that the fraction of zeros in $M_n(l)$ depends on the Hamming weight $w_l$ of $l$. For some fixed value of $n$, $w_l$ can assume $n$ different values namely $1,2,\cdots,n$ as $l$ varies between $1$ and $2^n-1$. Determination of $w_l$ for which the fraction of zeros in $M_n^{(l)}$ is minimum remains an unsolved problem. In Table \ref{tab1}, the variation of fraction of zeros with the Hamming weight of $l$ is illustrated for $n=4,5,6,7$. Observe that for $n=6$ the fraction of zeros when $l=2$ is lower than when $l=1.$

\section{Simulation Results}
\label{sec4}
\begin{figure}[p]
\centerline{\psfig{figure=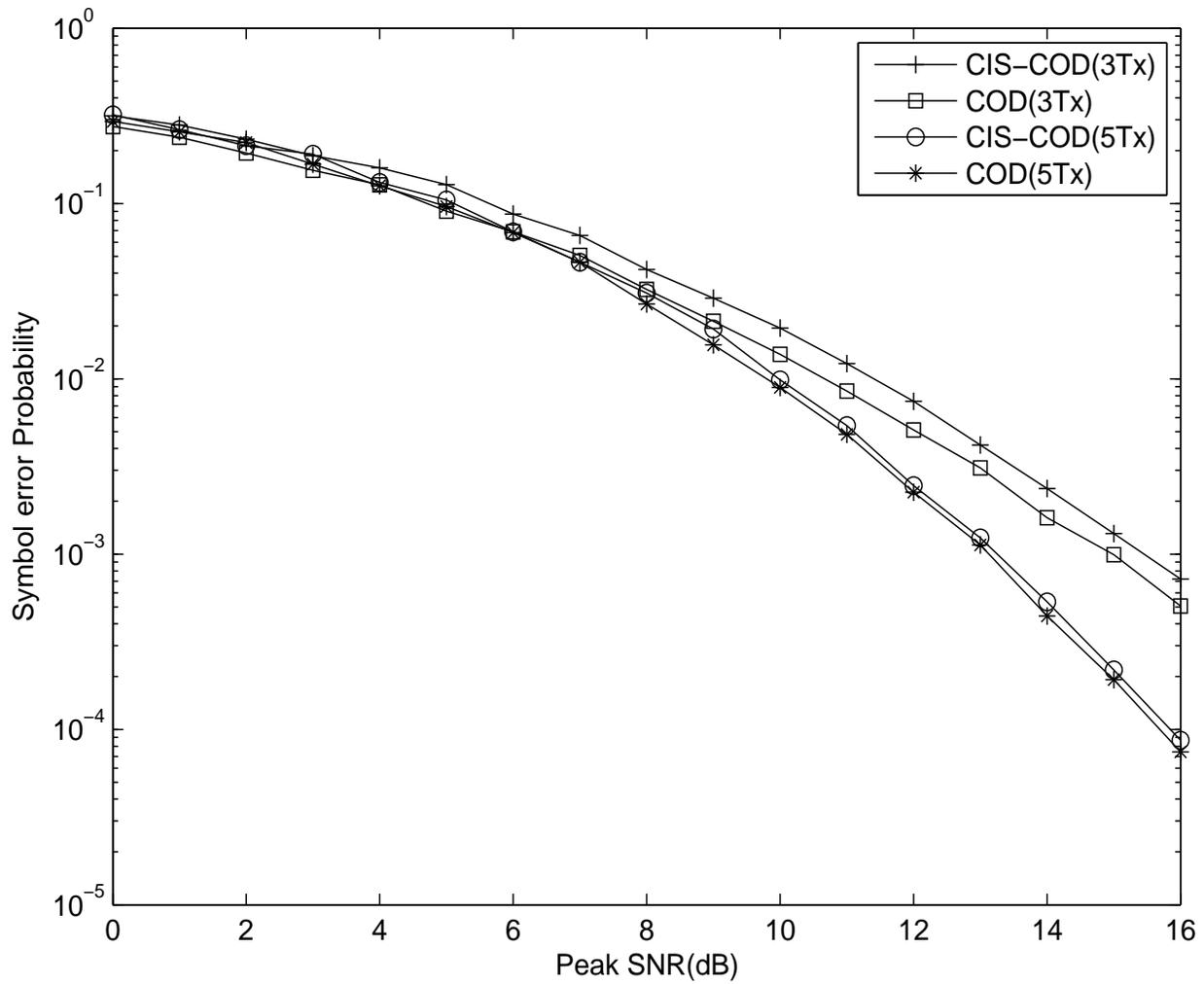,height=5.8in,width=7.4in}}
\caption{The performance of the CODs for 3 and 5 Tx. antennas using QAM modulation under peak power constraint.} \label{c1fig1}
\end{figure}

\begin{figure}[p]
\centerline{\psfig{figure=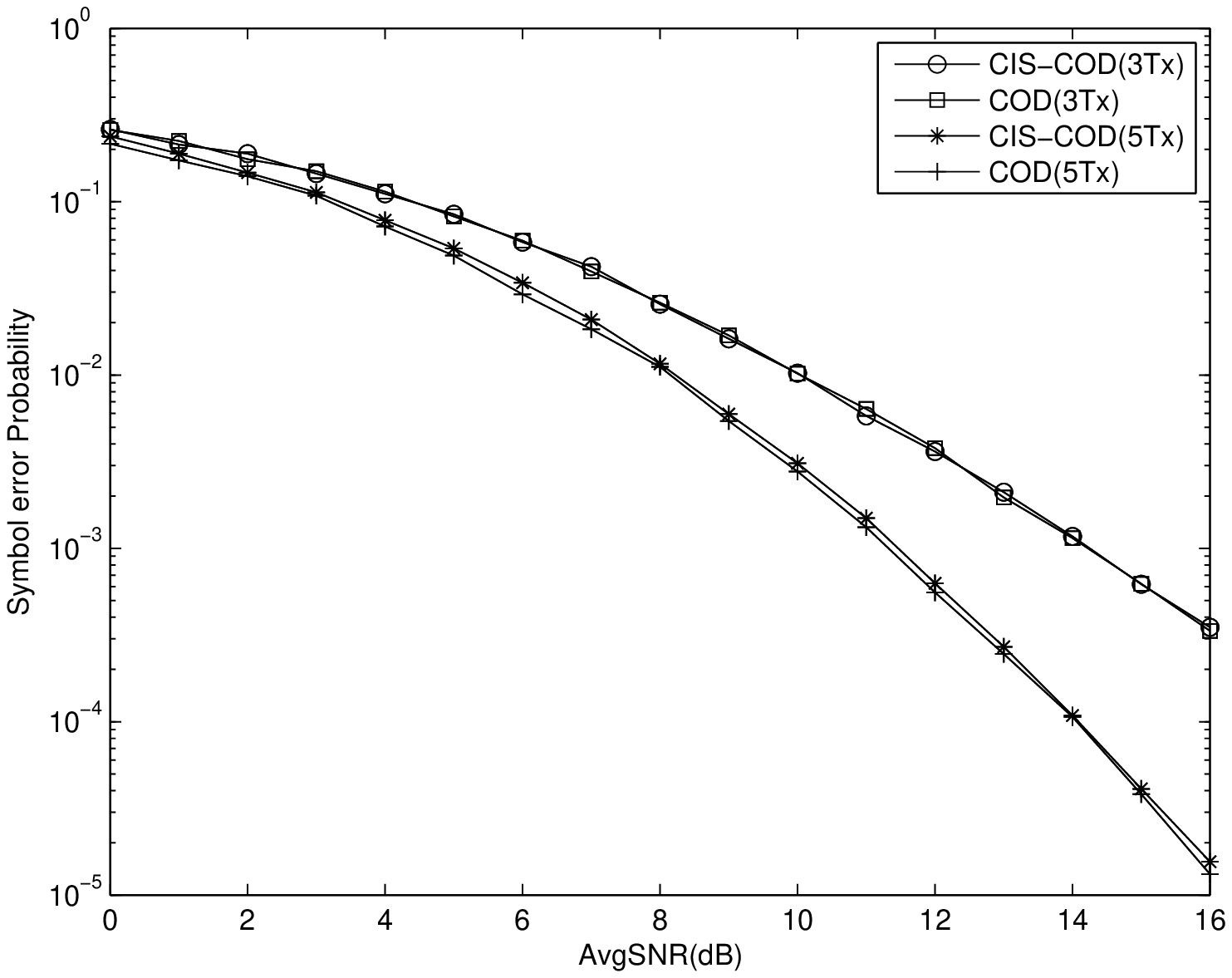,height=5.8in,width=7.4in}}
\caption{The performance of the CODs for 3 and 5 Tx. antennas using QAM modulation under average power constraint.} \label{c1fig2}
\end{figure}

The symbol error performance of the maximal rate CODs $L_4$ and $L_5$ for $3$ and $5$ transmit antennas with fewer number of zeros constructed in this paper (denoted as CIS-COD in Fig. \ref{c1fig1} and in Fig. \ref{c1fig2}) are compared with the existing maximal rate codes (denoted as COD) for same number of antennas in Fig. \ref{c1fig1} under peak power constraint. It is seen that in both the cases CIS-COD perform better than the CODs. 
Similarly, Fig. \ref{c1fig2} gives performance comparison of the corresponding codes under average power constraint. It is clear that the performance under average power constraint of the CIS-CODs  match with that of the corresponding CODs.  
\section{Discussion}
\label{sec5}

We have constructed a class of  maximal-rate $\frac{t+1}{2t}$ rectangular CIS-CODs for $2t-1$ or $2t$ transmit antennas, for all values of $t$, with lesser number of zero entries than the known constructions. Along the way, we have also devised a method of construction of maximal rate achievable CODs as given in \cite{Lia,LFX} from square CODs which can be viewed as the generalization of the method of construction of rate- $1$ RODs from square RODs for complex orthogonal designs. For the number of antennas $n$ our class of new codes has $n-1$ CIS-CODs indexed by $\norm{l}=1,2,\cdots, n-1,$  with the PAPR depending on $\norm{l}.$ An important direction for further research is to identify $l$ for which the PAPR is minimum.

\begin{appendices}
\section{Proof of Theorem \ref{thm3}} 
\label{append1}
We use Lemma~\ref{propercomplex} to prove this theorem. As the map $g: Z_{k_{a+1}}\rightarrow C_a$ is bijective, $g^{-1}$ is also bijective and hence $\lambda_{H^\prime_a}$ is injective if one of its arguments is kept fixed. Therefore, each variable appears exactly once in each column and at most once in each row of $H^\prime_a$.
Secondly, assuming that $H^\prime_a(i,j),H^\prime_a(i^\prime,j^\prime),H^\prime_a(i,j^\prime)$ are non-zero and
$\vert H^\prime_a(i,j)\vert=\vert H^\prime_a(i^\prime,j^\prime)\vert$ for some $i,i^\prime\in Z_{p_{a+1}},j,j^\prime\in Z_{k_{a+1}}$, we show that  $H^\prime_a(i^\prime,j)$ is non-zero and
$\vert H^\prime_a(i^\prime,j)\vert=\vert H^\prime_a(i,j^\prime)\vert$ as follows: since $\vert H^\prime_a(i,j)\vert=\lambda_{H^\prime_a}$, we have $f(i)\oplus  2_+^{j-1}=f(i^\prime)\oplus 2_+^{j^\prime-1}$. 

Next we show that any proper $2\times 2$ sub-matrix of $H^\prime_a$ is a COD of size $[2,2,2]$.
Note that $H^\prime_a(\alpha,\beta)\neq 0$ if and only if $f(\alpha)\in 2_+^{\beta-1}\oplus C_a$.

Let $M_2$ be a proper $2\times 2$ sub-matrix of $H^\prime_a$ formed by two distinct rows namely $i$ and $j$ and two distinct columns, say, $k$ and $l$. 
Then the entries of $M_2$ are given by $H^\prime_a(i,k)$, $H^\prime_a(i,l)$, $H^\prime_a(j,k)$ and $H^\prime_a(j,l)$. It is clear that $i\neq j$ and $k\neq l$. We always assume $l$ to be non-zero.\\
We show that $M_2$ is of the form
 $ \left[\begin{array}{cc}
    ay_\alpha   & by_\beta      \\
    cy_\beta^* & dy_\alpha^*
    \end{array}\right] \quad
or \quad
  \left[\begin{array}{cc}
  ay_\alpha^*   & by_\beta^*      \\
  cy_\beta     & dy_\alpha
  \end{array}\right]$,
with $\alpha,\beta\in Z_{k_{a+1}}$, $a,b,c,d\in\{1,-1\}$ satisfying $abcd=-1$.
In other words, we have to prove that\\
(A) $\mu(i,k)\mu(i,l)\mu(j,k)\mu(j,l)=-1$,~~(B) $\tau(i,k)\tau(j,k)=-1$ and (C) $\tau(i,k)\tau(i,l)=1$ where we write $\mu$ and $\tau$ in stead of $\mu_{H^\prime_a}$ and
$\tau_{H^\prime_a}$ respectively.
As all the entries of $M_2$ are non-zero, we have
\[
\{f(i)\oplus 2_+^{k-1}, f(i)\oplus 2_+^{l-1}, f(j)\oplus 2_+^{k-1}, f(j)\oplus 2_+^{l-1}\}\subset C_a.
\]
As $M_2$ is a proper sub-matrix of $H_a^{\prime}$, it contains only two distinct variables which implies that $\lambda(i,k)=\lambda(j,l)$ and $\lambda(i,l)=\lambda(j,k)$. i.e., $g^{-1}(f(i)\oplus 2_+^{k-1})=g^{-1}(f(j)\oplus 2_+^{l-1})$.
As $g$ is bijective, we have
\begin{equation}
\label{fifj}
f(i)\oplus f(j)= 2_+^{k-1}\oplus 2_+^{l-1}.
\end{equation}
(A) We first show that $\mu(i,k)\mu(i,l)\mu(j,k)\mu(j,l)=-1$.\\
{\bf Case (i)  $k=0:$}
We have $\mu(i,k)=1$, $\mu(j,k)=1$ and $f(i)\oplus f(j)=2^{l-1}$.
Hence one has to show that $\mu(i,l)\mu(j,l)=-1$ i.e., $(-1)^{1+ \norm{f(i)\cdot \overline{2^{l-1}}}+1+ \norm{ f(j)\cdot \overline{2^{l-1}}}}=-1$.
But $\norm{( f(i)\oplus f(j))\cdot \overline{2^{l-1}}}$ is an odd number
as $\norm{2^{l-1}\cdot \overline{2^{l-1}}}$ is an odd number.\\
{\bf Case (ii)  $k\neq 0:$}
It is enough to show that
$\norm{(f(i)\oplus f(j))\cdot (\overline{2^{k-1}}\oplus \overline{2^{l-1}})}$ is an odd number.
But $f(i)\oplus f(j)=2^{k-1}\oplus 2^{l-1}$ and
$\norm{(2^{k-1}\oplus 2^{l-1})\cdot (\overline{2^{k-1}}\oplus \overline{2^{l-1}})}$
is an odd number.\\

(B) We now prove that $\tau(i,k)\tau(j,k)=-1$.\\
{\bf Case (i)  $k=0:$}
From \eqref{fifj}, we have $f(i)\oplus f(j)=2^{l-1}$ and hence
$\norm{f(i)}+\norm{f(j)}$ 
is an odd number
as $f(i)\oplus f(j)= 2^{l-1}$ for some $l$.\\
{\bf Case (ii)  $k\neq 0:$}
We have
\begin{eqnarray*}
\norm{f(i)\cdot 2^{k-1}} +\norm{f(j)\cdot 2^{k-1}}
\equiv \norm{(f(i) \oplus f(j))\cdot 2^{k-1}}
=\norm{(2^{k-1}\oplus 2^{l-1})\cdot 2^{k-1}}=1.
\end{eqnarray*}

(C) Finally, we show that $\tau(i,k)\tau(i,l)=1$.\\
For $k=0$, $\norm{f(i)}+\lceil\frac{a}{2}\rceil +1 +\norm{f(i)\cdot 2^{l-1}}$
is an even number by Lemma~\ref{alp} as $f(i)\in C_a\cap (2^{l-1}\oplus C_a)$.
For $k\neq 0$, we have
\[
\tau(i,k)\tau(i,l)=1 +\norm{f(i)\cdot 2^{k-1}}+1 +\norm{f(i)\cdot 2^{l-1}}.
\]
Now $f(i)\in 2^{k-1}\oplus C_a$ and $f(i)\in 2^{l-1}\oplus C_a$ i.e., $(f(i)\oplus 2^{k-1})\in C_a \cap (2^{k-1}\oplus 2^{l-1}\oplus C_a)$.\\
Hence, by Lemma~\ref{alp}, $\norm{(f(i)\oplus 2^{k-1})\cdot 2^{k-1}} +\norm{(f(i)\oplus 2^{k-1})\cdot 2^{l-1}}=1$. \\
Therefore, $\norm{f(i)\cdot (2^{k-1}\oplus 2^{l-1})}$ is an even number. This concludes the proof. 

\section{Proof of Theorem \ref{thm4}}
\label{append2}
It is enough to show that the $(4m-1)$-th column of $\hat{H}_{4m}$ is orthogonal to all other columns of the matrix.
We use Lemma~\ref{propercomplex} to prove this statement.
All the complex variables appear exactly once in the $(4m-1)$-th column of $\hat{H}_{4m}$ which follows from \eqref{Hhat4m1}.
Secondly, assuming that $\hat{H}_{4m}(i,j),H^\prime_a(i^\prime,4m-1),H^\prime_a(i,4m-1)$ are non-zero and
$\vert\hat{H}_{4m}(i,j)\vert=\vert \hat{H}_{4m}(i^\prime,4m-1)\vert$ for some $i,i^\prime\in Z_{p_{a+1}},j\in Z_{k_{a+1}}$, we show that  $\hat{H}_{4m}(i^\prime,j)$ is non-zero and
$\vert \hat{H}_{4m}(i^\prime,j)\vert=\vert \hat{H}_{4m}(i,4m-1)\vert$.
But $\vert \hat{H}_{4m}(i,j)\vert=\lambda_{\hat{H}_{4m}}(i,j)$, hence $f(i)\oplus 2_+^{j-1}=f(i^\prime)\oplus \mathbf{1}$.

Let $M_2$ be a proper $2\times 2$ sub-matrix of $\hat{H}_{4m}$ formed by two distinct rows namely $i$ and $j$ and two distinct columns, say, $k$ and $l$ where $l$ is equal to $4m-1$. 
Then the entries of $M_2$ are given by $\hat{H}_{4m}(i,k)$, $\hat{H}_{4m}(i,l)$, $\hat{H}_{4m}(j,k)$ and $\hat{H}_{4m}(j,l)$.\\
We show that $M_2$ is of the form

$ \left[\begin{array}{cc}
    ay_\alpha   & by_\beta^*      \\
    cy_\beta    & dy_\alpha^*
    \end{array}\right]$
with $\alpha,\beta\in Z_{k_{a+1}}$, $a,b,c,d\in\{1,-1\}$ satisfying $abcd=-1$.
In other words, it is enough to prove that\\
(A) $\mu(i,k)\mu(i,l)\mu(j,k)\mu(j,l)=-1$,~~
(B) $\tau(i,k)\tau(j,k)=1$,~~
(C) $\tau(i,k)\tau(i,l)=-1$.\\
In order to have all the entries of $M_2$ non-zero, one must have
\begin{equation}
\label{subsetofc}
\{f(i)\oplus 2_+^{k-1}, f(i)\oplus \mathbf{1}, f(j)\oplus 2_+^{k-1}, f(j)\oplus \mathbf{1}\}\subset C_{4m-2}.
\end{equation}
As $M_2$ is a proper sub-matrix of $\hat{H}_{4m}$, it contains only two distinct variables which implies that $\lambda(i,k)=\lambda(j,l)$ and $\lambda(i,l)=\lambda(j,k)$. i.e., $g^{-1}(f(i)\oplus 2_+^{k-1})=g^{-1}(f(j)\oplus \mathbf{1})$.
As $g$ is bijective, we have
\begin{equation}
\label{fifj1}
f(i)\oplus f(j)= 2_+^{k-1}\oplus \mathbf{1}.
\end{equation}
To prove (A), we have following two cases.\\
{\bf Case (i)  $k=0:$}
It is enough to prove that $\mu(i,l)\mu(j,l)=-1$ i.e., $(-1)^{1+\norm{f(i)\cdot \hat{\mathbf{1}}}}\cdot (-1)^{1+\norm{f(j)\cdot \hat{\mathbf{1}}}}=-1$.
By \eqref{fifj1}, $f(i)\oplus f(j)=\mathbf{1}$ and $\norm{\mathbf{1}\cdot \hat{\mathbf{1}}}$ is an odd number.\\
{\bf Case (ii)  $k\neq 0:$}
It is enough to show that
$\norm{(f(i)\oplus f(j))\cdot (\overline{2^{k-1}}\oplus \hat{\mathbf{1}})}$
i.e.,
$\norm{(2^{k-1}\oplus \mathbf{1})\cdot (\overline{2^{k-1}}\oplus \hat{\mathbf{1}})}$ is an odd number.
As both $\norm{2^{k-1}\cdot \overline{2^{k-1}}}$ and $\norm{\mathbf{1}\cdot \hat{\mathbf{1}}}$ are odd numbers, $\norm{2^{k-1} \cdot \hat{\mathbf{1}}}+\norm{\mathbf{1}\cdot \overline{2^{k-1}}}$ must be an odd number. This is indeed true as
$\norm{2^{k-1}.\hat{\mathbf{1}}}$ is $1$ or $0$ respectively if $(k-1)$ is even or odd
and $\norm{\overline{2^{k-1}}.\mathbf{1}} =4l+2-(k-1)$.\\
We now prove the statement (B) i.e., $\tau(i,k)\tau(j,k)=1$.\\
{\bf Case (i)  $k=0:$}
We have $f(i)\oplus f(j)=\mathbf{1}$ and
$\norm{f(i)}+\norm{f(j)}\equiv\norm{\mathbf{1}}$
is an even number. \\
{\bf Case (ii)  $k\neq 0:$}
If $k=4m-1$, we have $\tau(i,k)=-1$ for all $i$ whenever $\hat{H}_{4m}(i,k)\neq 0$ which implies that $\tau(i,k)\tau(j,k)=1$.\\
If $k< 4m-1$, we have
\begin{eqnarray*}
\norm{f(i)\cdot 2^{k-1}} +\norm{f(j)\cdot 2^{k-1}}
\equiv \norm{(f(i) \oplus f(j))\cdot 2^{k-1}}
=\norm{(2^{k-1}\oplus \mathbf{1})\cdot 2^{k-1}}=0.
\end{eqnarray*}
(C) Finally, we show that $\tau(i,k)\tau(i,l)=-1$ with $l=4m-1$.\\
As $\tau(i,l)=-1$, one has to prove that $\tau(i,k)=1$ whenever
$f(i)\in (2_+^{k-1}\oplus C_{4m-2})\cap (\mathbf{1}+C_{4m-2})$.\\
If $k=0$, then $f(i)\in C_{4m-2}\cap (\mathbf{1}+C_{4m-2})$.\\
By \eqref{tk}, $C_{4m-2}=\Big\{w\in Z_{2^{4m-2}}~\Big\vert~ \norm{w}= 2m-2 \text{ or } 2m-1\Big\}$.\\
Hence $C_{4m-2}\cap \mathbf{1}+C_{4m-2}=\Big\{w\in Z_{2^{4m-2}}~\Big\vert~ \norm{i}=2m-1\Big\}$.
Therefore, $\norm{f(i)}+\lceil\frac{4m-2}{2}\rceil$ is an even number.\\
If $k\neq 0$, it is enough to prove that $\norm{f(i)\cdot 2^{k-1}}=1$
for all $f(i)\in (2^{k-1}\oplus C_{4m-2})\cap (\mathbf{1}+C_{4m-2})$.\\
As $f(i)\in \mathbf{1}\oplus C_{4m-2}$, we have $\norm{f(i)}=2m-1 \mbox{ or } 2m$.
Again $f(i)\oplus 2^{k-1}\in C_a$ which implies that $\norm{f(i)\oplus 2^{k-1}}=2m-2 \mbox{ or } 2m-1$.
Now
\[
\norm{f(i)\oplus 2^{k-1}}=\norm{f(i)}+\norm{2^{k-1}} -2\norm{f(i)\cdot 2^{k-1}}.
\]
As $k\neq 0$, we have $\norm{2^{k-1}}=1$. Thus $\norm{f(i)\oplus 2^{k-1}}$ and $\norm{ f(i)}$ differ by an odd number.\\
So $(\norm{f(i)}, \norm{f(i)+2^{k-1}})=(2m,2m-1)$ or $(2m-1,2m-2)$.
In both cases, $\norm{f(i)\cdot 2^{k-1}}=1$.
This concludes the proof. 
\section{Proof of Theorem \ref{thm5}}
\label{append3}
We give an explicit construction of the code for any number of transmit antennas with the specified rate and fraction of zeros. This matrix is obtained from $H^\prime_n$ after performing row operations defined in \eqref{row_oper}. This code is denoted by $M_n(l)$. The rate of this code matches with that of $H^\prime_n$.
We now show that the fraction of zeros in $M_n(l)$ is as given in the statement of the theorem.  Let
$Pp_{n}$ be the set of complex variables which form a pair with another variable. Similarly,
${Up}_{n},Pr_{n}$ and $Ur_{n}$ respectively denote the set of complex variables which don't form a pair, the set of rows which form a pair with another row and the set of rows which don't form a pair.
It is assumed that the elements of the sets are not the rows or the variables themselves
denoted by $R_i,~i=0,1,\cdots$ and $y_i,~i=0,1,\cdots$ respectively, but the indices of the rows or the variables i.e., the numbers $1,2,3,\cdots$. By definition, we have
\begin{eqnarray*}
Z_{k_{n+1}}=Pp_{n}\cup Up_{n},~ Z_{p_{n+1}}=Pr_{n}\cup Ur_{n},~~~~ Pp_{n}\cap Up_{n}=\phi, ~Pr_{n}\cap Ur_{n}=\phi,\\
Pp_{n}=\{i\mid g(i)\in C_{n}\cap (l\oplus C_{n})\},~~~~Pr_{n}=\{i\mid f(i)\in R_{n}\cap (l\oplus R_{n})\}.
\end{eqnarray*}
Therefore, $g(Pp_n)=C_n\cap (l\oplus C_n), f(Pr_n)=R_n\cap (l\oplus R_n)$. \\
We first compute  the number of variables that appear twice in a column of the matrix  $M_{n}$. It need not be true that this value is same for all columns of $M_n(l)$.
We fix a column of the matrix $M_n(l)$, say $m$ where $m\in\{0,1,\cdots,n\}$.
Let $W_n^{(l,m)}\subset Z_{k_{n+1}}$ be the set of variables that appear twice in $m$-th column of the matrix $M_{n}$ and $e_{n+1}^{(l,m)}=\vert W_{n}^{(l,m)}\vert$.\\  
The number of non-zero entries in $m$-th column is $k_{n+1}+e_{n+1}^{(l,m)}$ where $k_{n+1}$ is the number of complex variables in the COD $H^\prime_n$. We have $F_{n+1}(l)=1-\frac{\sum_{m=0}^{n}(k_{n+1}+e_{n+1}^{(l,m)})}{ (n+1)\cdot p_{n+1}}$. Assigning $e_{n+1}^{(l)}=\frac{\sum_{m=0}^{n}e_{n+1}^{(l,m)}}{n+1}$, we get
\begin{eqnarray}
\label{fznp1}
F_{n+1}(l)=1-\frac{(k_{n+1}+e_{n+1}^{(l)})}{p_{n+1}}.
\end{eqnarray}
In the following, we calculate $e_{n+1}^{(l)}$.
Let $C_n^{(m)}= 2_+^{m-1}\oplus C_{n}$. Note that $C_n^{(m)}\subset R_n$. We have
{\footnotesize
\begin{eqnarray*}
2_+^{m-1}\oplus g(W_n^{(l,m)}) &=&(2_+^{m-1}\oplus g(Up_n))\cap f(Pr_n)\\
&=&(2_+^{m-1}\oplus (C_n\setminus(l\oplus C_n)))\cap (R_n\cap (l\oplus R_n))\\
&=&((2_+^{m-1}\oplus C_n)\setminus (2^{m-1}\oplus l\oplus C_n))\cap (R_n\cap (l\oplus R_n))\\
&=&(C_n^{(m)}\setminus(l\oplus C_n^{(m)})) \cap (l\oplus R_n)\\
&=&C_n^{(m)}\cap (l\oplus (R_{n}\setminus C_n^{(m)})).\\
\mbox{ For $m \neq 0,$ } \hspace{2.00cm} C_n^{(m)}&=&\{i\in Z_{2^n}\mid i_{m-1}=1 \text { and } \norm{i}=\Big\lceil \frac{n}{2}\Big\rceil
 \text{ or } \norm{i}=\Big\lceil \frac{n}{2}\Big\rceil+1\} \\
&&\cup \{i\in Z_{2^n}\mid i_{m-1}=0 \text { and } \norm{i}=\Big\lceil \frac{n}{2}\Big\rceil-2 \text{ or } \norm{i}=\Big\lceil \frac{n}{2}\Big\rceil-1\}, \mbox{   and   }\\
 R_{n}\setminus C_n^{(m)}&=&\{i\in Z_{2^n}\mid i_{m-1}=0 \text { and } \norm{i}=\Big\lceil \frac{n}{2}\Big\rceil \text{ or } \norm{i}=\Big\lceil \frac{n}{2}\Big\rceil+1\} \\
&&\cup \{i\in Z_{2^n}\mid i_{m-1}=1 \text { and } \norm{i}=\Big\lceil \frac{n}{2}\Big\rceil-2 \text{ or } \norm{i}=\Big\lceil \frac{n}{2}\Big\rceil-1\}.
\end{eqnarray*}
}
For fixed $l>0$, and $m>0$, we have following four cases:
\begin{eqnarray*}
\begin{array}{rr}
(i)~ l_{m-1}=0,~ i_{m-1}=0  & (iii)~ l_{m-1}=1,~ i_{m-1}=0, \\
(ii)~ l_{m-1}=0,~ i_{m-1}=1  & (iv)~ l_{m-1}=1,~ i_{m-1}=1.
\end{array}
\end{eqnarray*}
\noindent
Let $t_1,t_2\in\{0,1\}$.
Define $Z_{n}^{(t_1,t_2)}=\{i\in C_n^{(m)}\cap (l\oplus R_{n}\setminus C_n^{(m)})\mid l_{m-1}=t_1,i_{m-1}=t_2\}$.
For case (1), we have

{\footnotesize
\begin{eqnarray*}
Z_n^{(0,0)}&&=\{i\in Z_{2^n}\mid~i_{m-1}=0,\Big( \norm{i}=\Big\lceil \frac{n}{2}\Big\rceil-2  \text{ or }\Big\lceil \frac{n}{2}\Big\rceil-1\Big) \text { and } \Big(\norm{i\oplus l}=\Big\lceil \frac{n}{2}\Big\rceil \text{ or } \Big\lceil \frac{n}{2}\Big\rceil+1\Big)\}\\
&&=\{i\in Z_{2^n}\mid \Big(i_{m-1}=0, \norm{i}=\Big\lceil \frac{n}{2}\Big\rceil-2 \text{ and } \norm{l.i}=\Big\lceil \frac{ w_l-1}{2}\Big\rceil-1\Big)
 \text{ or }\\
&&~\Big(i_{m-1}=0, \norm{i}=\Big\lceil \frac{n}{2}\Big\rceil-1\text{ and } \norm{l.i}=\Big\lceil \frac{ w_l}{2}\Big\rceil-1\Big)\} \mbox{ where } w_l=\norm{l}, \mbox { the Hamming weight of $l$. }\\
\end{eqnarray*}

\begin{eqnarray*}
\vert Z_n^{(0,0)}\vert&&=\binom{w_l}{\Big\lceil\frac{w_l-1}{2}\Big\rceil-1} \cdot\binom{n-1-w_l)}{\Big\lceil \frac{n}{2}\Big\rceil-2-\Big\lceil\frac{ w_l-1}{2}\Big\rceil+1} +
\binom{w_l}{\Big\lceil\frac{w_l}{2}\Big\rceil-1} \cdot\binom{n-1-w_l)}{\Big\lceil \frac{n}{2} \Big\rceil-1-\Big\lceil\frac{ w_l}{2}\Big\rceil+1}\\
&&=\binom{w_l}{\Big\lceil\frac{w_l-1}{2}\Big\rceil-1} \cdot\binom{n-w_l-1)}{\Big\lceil \frac{n}{2}\Big\rceil-\Big\lceil\frac{ w_l-1}{2}\Big\rceil-1}+
\binom{w_l}{\Big\lceil\frac{w_l}{2}\Big\rceil-1} \cdot\binom{n-w_l-1)}{\Big\lceil \frac{n}{2} \Big\rceil-\Big\lceil\frac{ w_l}{2}\Big\rceil}\\
&&=\binom{2\Big\lceil\frac{w_l}{2}\Big\rceil}{\Big\lceil\frac{w_l}{2}\Big\rceil-1} \cdot\binom{n-2\Big\lceil\frac{w_l}{2}\Big\rceil)}{\Big\lceil \frac{n}{2} \Big\rceil-\Big\lceil\frac{ w_l}{2}\Big\rceil}.
\end{eqnarray*}
}
Similarly, we have

{\footnotesize
\begin{eqnarray*}
\vert Z_n^{(0,1)}\vert=\binom{2\Big\lceil\frac{w_l}{2}\Big\rceil}{\Big\lceil
\frac{w_l}{2}\Big\rceil+1} \cdot\binom{n-2\Big\lceil\frac{w_l}{2}\Big\rceil}{\Big\lceil \frac{n}{2} \Big\rceil-\Big\lceil\frac{ w_l}{2}\Big\rceil-1},
\vert Z_n^{(1,0)}\vert=\binom{2\Big\lceil\frac{w_l}{2}\Big\rceil-1}{\Big\lceil\frac{w_l}{2}
\Big\rceil} \cdot\binom{n-2\Big\lceil\frac{w_l}{2}\Big\rceil+1)}{\Big\lceil \frac{n}{2} \Big\rceil-\Big\lceil\frac{ w_l}{2}\Big\rceil-1}, \\
\vert Z_n^{(1,1)}\vert =\binom{2\Big\lceil\frac{w_l}{2}\Big\rceil-1}{\Big\lceil\frac{w_l}{2}\Big\rceil-1} \cdot\binom{n-2\Big\lceil\frac{w_l}{2}\Big\rceil+1)}{\Big\lceil \frac{n}{2} \Big\rceil-\Big\lceil\frac{ w_l}{2}\Big\rceil+1}.
\end{eqnarray*}
}
Let $p_n^{(0)}=\vert Z_n^{(0,0)}\vert+\vert Z_n^{(0,1)}\vert$
and $p_n^{(1)}=\vert Z_n^{(1,0)}\vert+\vert Z_n^{(1,1)}\vert$. We have

{\footnotesize
\begin{eqnarray*}
p_n^{(0)} =\binom{2\Big\lceil\frac{w_l}{2}\Big\rceil}
{\Big\lceil\frac{w_l}{2}\Big\rceil+1} \cdot\binom{n-2\Big\lceil\frac{w_l}{2}\Big\rceil+1}{\Big\lceil \frac{n}{2} \Big\rceil-\Big\lceil\frac{ w_l}{2}\Big\rceil}, ~~~~~~~~~
p_n^{(1)}=
\begin{cases}
 \binom{2\Big\lceil\frac{w_l}{2}\Big\rceil}{\Big\lceil\frac{w_l}{2}\Big\rceil}
\cdot\binom{n-2\Big\lceil\frac{w_l}{2}\Big\rceil+1)}{ \frac{n+1}{2}-\Big\lceil\frac{ w_l}{2}\Big\rceil+1}  ~~~~ \text{ if $n$ is odd, }\\
    \binom{2\Big\lceil\frac{w_l}{2}\Big\rceil-1}{\Big\lceil\frac{w_l}{2}\Big\rceil}
\cdot\binom{n-2\Big\lceil\frac{w_l}{2}\Big\rceil+2)}{\frac{n}{2} -\Big\lceil\frac{ w_l}{2}\Big\rceil+2} ~~~~ \text{ if $n$ is even. }
\end{cases} \\
\end{eqnarray*}
\begin{eqnarray*}
\mbox{ For $m \neq 0, ~~~~~~~~~~~~$}   \vert W_n^{(l,m)}\vert=
    \begin{cases}
p_n^{(0)}    & \text{ if $l_{m-1}=0$ }\\
p_n^{(1)}    & \text{ if $l_{m-1}=1$ }.
  \end{cases} \\
\end{eqnarray*}
\begin{eqnarray*}
\mbox{ For $m=0$ }  \hspace{1.00cm} g(W_n^{(l,0)})&&=C_n\cap (l\oplus (R_n\setminus C_n))\\
\vert W_{n}^{(l,0)} \vert&&=
 \binom{w_l}{\Big\lceil\frac{w_l+1}{2}\Big\rceil} \left(\binom{n-w_l}{ \Big\lceil\frac{n}{2}\Big\rceil-\Big\lceil\frac{ w_l}{2}\Big\rceil-1}+
\binom{n-w_l}{ \Big\lceil\frac{n}{2}\Big\rceil-\Big\lceil\frac{ w_l+1}{2}\Big\rceil+1}\right)\\&&=   \begin{cases}
\binom{w_l}{\Big\lceil\frac{w_l+1}{2}\Big\rceil}\cdot\binom{n-w_l+1}{ \frac{n}{2}-\Big\lceil\frac{ w_l}{2}\Big\rceil} & \text{ if $n$ is even, } \nonumber\\
2\binom{w_l}{\Big\lceil\frac{w_l+1}{2}\Big\rceil}\cdot\binom{n-w_l}{ \frac{n+1}{2}-\Big\lceil\frac{ w_l+1}{2}\Big\rceil+1} & \text{ if $n$ is odd. } \nonumber\\
\end{cases}
\end{eqnarray*}
}
When $n$ is odd and $w_l=n$, we observe that $g(W_n^{(l,m)})$ is a null set, hence $\vert W_{n}^{(l,m)}\vert=0$ for $m=0,1,\cdots,n$. We can therefore assume that $\binom{0}{1}=0$, so that $\vert W_{n}^{(l,0)}\vert=v_n^{(1)}$ is zero  and
the fraction of zeros in $M_n(l)$ is equal to $\frac{1}{2}-\frac{1}{n+1}$ in this case.
In all other cases, we get the required expression for the fraction of zeros by using \eqref{fznp1}. 
This completes the proof.
\end{appendices}
\onecolumn

\onecolumn

\begin{table*}
\caption{Variation of fraction of zeros in $M_n(l)$ with the Hamming weight $w_l$ of $l,$ for $n=4,5,6,7$}
\begin{center}
\begin{tabular}{|c|l|l|l|l|l|l|l|l|} \hline  \label{tab1}
$w_l$             & 1 & 2& 3& 4 & 5 & 6& 7   \\ \hline 
Fraction of zeros in $M_4(l)$ & 0.1067   & 0.1067   & 0.1200   & 0.1200&&& \\ \hline
Fraction of zeros in $M_5(l)$ & 0.1111  &  0.1111  &  0.1111 &   0.1111 & .3333&& \\ \hline
Fraction of zeros in $M_6(l)$ & 0.1709   &0.1709   &0.1607   &0.1607   & 0.1837   &0.1837& \\ \hline
Fraction of zeros in $M_7(l)$ & 0.1741   &0.1741   &0.1607   &0.1607   & 0.1741 &0.1741 & .3750  \\ \hline
\end{tabular}
\end{center}
\end{table*}

\newpage

\onecolumn
\begin{figure*}
\begin{eqnarray}
\label{h4h4}
\widetilde{H}_4&=&\left[\begin{array}{rrrrrrrrrr}
            -x_1^* &-x_2^* &0    &-x_3^*   &0      &0     &-x_4^* &0       &0      &0
\\
             x_0^*  &0    &-x_2^*  &0     &-x_3^*  &0      &0    &-x_4^*   &0      &0
\\
             0      &x_0^* &x_1^*  &0      &0     &-x_3^*  &0      &0     &-x_4^*  &0
\\
             x_2    &-x_1  &x_0    &0      &0      &0      &0      &0      &0      &0
\\
             0      &0     &0     &x_0^*   &x_1^*  &x_2^*  &0      &0      &0     &-x_4^*\\
             x_3    &0     &0    &-x_1     &x_0    &0      &0      &0      &0      &0
\\
             0      &x_3   &0    &-x_2     &0      &x_0    &0      &0      &0      &0
\\
             0      &0     &x_3    &0     &-x_2    &x_1    &0      &0      &0      &0
\\
             0      &0     &0      &0      &0      &0      &x_0^*  &x_1^*  &x_2^* &x_3^*\\
             x_4    &0     &0      &0      &0      &0     &-x_1    &x_0    &0      &0
\\
             0      &x_4   &0      &0      &0      &0     &-x_2    &0      &x_0    &0
\\
             0      &0     &x_4    &0      &0      &0      &0     &-x_2    &x_1    &0
\\
             0      &0     &0      &x_4    &0      &0     &-x_3    &0      &0      &x_0
\\
             0      &0     &0      &0      &x_4    &0      &0     &-x_3    &0      &x_1
\\
             0      &0     &0      &0      &0      &x_4    &0      &0     &-x_3    &x_2
 \end{array}\right],~~  \nonumber \\
H_4&=&\left[\begin{array}{rrrrr}
              0             &-y_0^*        &-y_1^*          &-y_3^*       &-y_6^*\\
              y_0^*         &0             &-y_2^*          &-y_4^*       &-y_7^*    \\
              y_1^*         &y_2^*         &0             &-y_5^*       &-y_8^*    \\
              y_2           &-y_1          &y_0              &0            &0    \\
              y_3^*         &y_4^*         &y_5^*              &0         &-y_9^* \\
              y_4           &-y_3          &0             &y_0            &0    \\
              y_5           &0             &-y_3             &y_1            &0    \\
              0             &y_5           &-y_4             &y_2             &0    \\
              y_6^*         &y_7^*         &y_8^*          &y_9^*          &0    \\
              y_7           &-y_6          &0              &0             &y_0    \\
              y_8           &0             &-y_6          &0             &y_1    \\
              0             &y_8           &-y_7          &0             &y_2    \\
              y_9           &0             &0             &-y_6             &y_3    \\
              0             &y_9           &0             &-y_7             &y_4    \\
              0             &0             &y_9           &-y_8             &y_5
\end{array}\right]
\end{eqnarray}
\caption{CODs $\widetilde{H}_4$ of size [15,10,5] and $H_4$ of size [15,5,10]}
\label{fig3}

\end{figure*}

\newpage

\begin{figure*}

{\scriptsize
\begin{equation*}
\hat{H}_8=\left[  
\begin{array}{rrrrrrrr}
   0   &   0   &  -y_0^*   &  -y_1^*   &  -y_4^*   & -y_{10}^*   & -y_{20}^*   &   0\\
\vspace{-0.12cm}
   0   &   y_0^*   &   0   &  -y_2^*   &  -y_5^*   & -y_{11}^*   & -y_{21}^*   &   0\\
\vspace{-0.12cm}
   y_0^*   &   0   &   0   &  -y_3^*   &  -y_6^*   & -y_{12}^*   & -y_{22}^*   &   0\\
\vspace{-0.12cm}
   0   &   y_1^*   &   y_2^*   &   0   &  -y_7^*   & -y_{13}^*   & -y_{23}^*   &   0\\
\vspace{-0.12cm}
   y_1^*   &   0   &   y_3^*   &   0   &  -y_8^*   & -y_{14}^*   & -y_{24}^*   &   0\\
\vspace{-0.12cm}
   y_2^*   &  -y_3^*   &   0   &   0   &  -y_9^*   & -y_{15}^*   & -y_{25}^*   &   0\\
\vspace{-0.12cm}
   y_3   &  y_2   & -y_1   &  y_0   &   0   &   0   &   0   & -y_{34}^*\\
\vspace{-0.12cm}
   0   &   y_4^*   &   y_5^*   &   y_7^*   &   0   & -y_{16}^*   & -y_{26}^*   &   0\\
\vspace{-0.12cm}
   y_4^*   &   0   &   y_6^*   &   y_8^*   &   0   & -y_{17}^*   & -y_{27}^*   &   0\\
\vspace{-0.12cm}
   y_5^*   &  -y_6^*   &   0   &   y_9^*   &   0   & -y_{18}^*   & -y_{28}^*   &   0\\
\vspace{-0.12cm}
   y_6   &  y_5   & -y_4   &   0   &  y_0   &   0   &   0   & y_{33}^*\\
\vspace{-0.12cm}
   y_7^*   &  -y_8^*   &  -y_9^*   &   0   &   0   & -y_{19}^*   & -y_{29}^*   &   0\\
\vspace{-0.12cm}
   y_8   &  y_7   &   0   & -y_4   &  y_1   &   0   &   0   & -y_{32}^*\\
\vspace{-0.12cm}
   y_9   &   0   &  y_7   & -y_5   &  y_2   &   0   &   0   & y_{31}^*\\
\vspace{-0.12cm}
   0   & -y_9   &  y_8   & -y_6   &  y_3   &   0   &   0   & -y_{30}^*\\
\vspace{-0.12cm}
   0  &  y_{10}^*   &  y_{11}^*   &  y_{13}^*   &  y_{16}^*   &   0   & -y_{30}^*   &   0\\
\vspace{-0.12cm}
   y_{10}^* &   0   &  y_{12}^*   &  y_{14}^*   &  y_{17}^*   &   0   & -y_{31}^*   &   0\\
\vspace{-0.12cm}
   y_{11}^* & -y_{12}^*   &   0   &  y_{15}^*   &  y_{18}^*   &   0   & -y_{32}^*   &   0\\
\vspace{-0.12cm}
   y_{12}   & y_{11}   &-y_{10}   &   0   &   0   &  y_0   &   0   & -y_{29}^*\\
\vspace{-0.12cm}
   y_{13}^* & -y_{14}^*   & -y_{15}^*   &   0   &  y_{19}^*   &   0   & -y_{33}^*   &   0\\
\vspace{-0.12cm}
   y_{14}   & y_{13}   &   0   &-y_{10}   &   0   &  y_1   &   0   & y_{28}^*\\
\vspace{-0.12cm}
   y_{15}   &   0   & y_{13}   &-y_{11}   &   0   &  y_2   &   0   & -y_{27}^*\\
\vspace{-0.12cm}
   0   &-y_{15}   & y_{14}   &-y_{12}   &   0   &  y_3   &   0   & y_{26}^*\\
\vspace{-0.12cm}
   y_{16}^* & -y_{17}^*   & -y_{18}^*   & -y_{19}^*   &   0   &   0   & -y_{34}^*   &   0\\
\vspace{-0.12cm}
   y_{17}   & y_{16}   &   0   &   0   &-y_{10}   &  y_4   &   0   & -y_{25}^*\\
\vspace{-0.12cm}
   y_{18}   &   0   & y_{16}   &   0   &-y_{11}   &  y_5   &   0   & y_{24}^*\\
\vspace{-0.12cm}
   0   &-y_{18}   & y_{17}   &   0   &-y_{12}   &  y_6   &   0   & -y_{23}^*\\
\vspace{-0.12cm}
   y_{19}   &   0   &   0   & y_{16}   &-y_{13}   &  y_7   &   0   & -y_{22}^*\\
\vspace{-0.12cm}
   0   &-y_{19}   &   0   & y_{17}   &-y_{14}   &  y_8   &   0   & y_{21}^*\\
\vspace{-0.12cm}
   0   &   0   &-y_{19}   & y_{18}   &-y_{15}   &  y_9   &   0   & -y_{20}^*\\
\vspace{-0.12cm}
   0   &  y_{20}^*&  y_{21}^*   &  y_{23}^*   &  y_{26}^*   &  y_{30}^*   &   0   &   0\\
\vspace{-0.12cm}
   y_{20}^* &   0   &  y_{22}^*   &  y_{24}^*   &  y_{27}^*   &  y_{31}^*   &   0   &   0\\
\vspace{-0.12cm}
   y_{21}^* & -y_{22}^*   &   0   &  y_{25}^*   &  y_{28}^*   &  y_{32}^*   &   0   &   0\\
\vspace{-0.12cm}
   y_{22}   & y_{21}   &-y_{20}   &   0   &   0   &   0   &  y_0   & y_{19}^*\\
\vspace{-0.12cm}
   y_{23}^* & -y_{24}^*   & -y_{25}^*   &   0   &  y_{29}^*   &  y_{33}^*   &   0   &   0\\
\vspace{-0.12cm}
   y_{24}   & y_{23}   &   0   &-y_{20}   &   0   &   0   &  y_1   & -y_{18}^*\\
\vspace{-0.12cm}
   y_{25}   &   0   & y_{23}   &-y_{21}   &   0   &   0   &  y_2   & y_{17}^*\\
\vspace{-0.12cm}
   0   &-y_{25}   & y_{24}   &-y_{22}   &   0   &   0   &  y_3   & -y_{16}^*\\
\vspace{-0.12cm}
   y_{26}^* & -y_{27}^*   & -y_{28}^*   & -y_{29}^*   &   0   &  y_{34}^*   &   0   &   0\\
\vspace{-0.12cm}
   y_{27}   & y_{26}   &   0   &   0   &-y_{20}   &   0   &  y_4   & y_{15}^*\\
\vspace{-0.12cm}
   y_{28}   &   0   & y_{26}   &   0   &-y_{21}   &   0   &  y_5   & -y_{14}^*\\
\vspace{-0.12cm}
\vspace{-0.12cm}
   0   &-y_{28}   & y_{27}   &   0   &-y_{22}   &   0   &  y_6   & y_{13}^*\\
\vspace{-0.12cm}
   y_{29}   &   0   &   0   & y_{26}   &-y_{23}   &   0   &  y_7   & y_{12}^*\\
\vspace{-0.12cm}
   0   &-y_{29}   &   0   & y_{27}   &-y_{24}   &   0   &  y_8   & -y_{11}^*\\
\vspace{-0.12cm}
   0   &   0   &-y_{29}   & y_{28}   &-y_{25}   &   0   &  y_9   & y_{10}^*\\
\vspace{-0.12cm}
   y_{30}^*& -y_{31}^*   & -y_{32}^*   & -y_{33}^*   & -y_{34}^*   &   0   &   0   &   0\\
\vspace{-0.12cm}
   y_{31}   & y_{30}   &   0   &   0   &   0   &-y_{20}   & y_{10}   &  -y_9^*\\
\vspace{-0.12cm}
   y_{32}   &   0   & y_{30}   &   0   &   0   &-y_{21}   & y_{11}   &  y_8^*\\
\vspace{-0.12cm}
   0   &-y_{32}   & y_{31}   &   0   &   0   &-y_{22}   & y_{12}   &  -y_7^*\\
\vspace{-0.12cm}
   y_{33}   &   0   &   0   & y_{30}   &   0   &-y_{23}   & y_{13}   &  -y_6^*\\
\vspace{-0.12cm}
   0   &-y_{33}   &   0   & y_{31}   &   0   &-y_{24}   & y_{14}   &  y_5^*\\
\vspace{-0.12cm}
   0   &   0   &-y_{33}   & y_{32}   &   0   &-y_{25}   & y_{15}   &  -y_4^*\\
\vspace{-0.12cm}
   y_{34}   &   0   &   0   &   0   & y_{30}   &-y_{26}   & y_{16}   &  y_3^*\\
\vspace{-0.12cm}
   0   &-y_{34}   &   0   &   0   & y_{31}   &-y_{27}   & y_{17}   &  -y_2^*\\
\vspace{-0.12cm}
   0   &   0   &-y_{34}   &   0   & y_{32}   &-y_{28}   & y_{18}   &  y_1^*\\
\vspace{-0.12cm}
   0   &   0   &   0   &-y_{34}   & y_{33}   &-y_{29}   & y_{19}   &  -y_0^*
\end{array}\right]
\end{equation*}
}
\caption{Rate-5/8 COD of size $[56,8,35]$}
\label{c1fig3}

\end{figure*}
\newpage
\onecolumn
\begin{figure*}

\begin{eqnarray*}
L_4=\left[\begin{array}{rrrr}
    x_0   &-x_1^*  &-\frac{x_2^*}{\sqrt{2}} &-\frac{x_2^*}{\sqrt{2}}\\
    x_1   & x_0^*  &-\frac{x_2^*}{\sqrt{2}} &\frac{x_2}{\sqrt{2}}\\
    \frac{x_2}{\sqrt{2}}   & \frac{x_2}{\sqrt{2}}    & x_{1I}-jx_{0Q} & x_{0I}+jx_{1Q}\\
    \frac{x_2}{\sqrt{2}}   &-\frac{x_2}{\sqrt{2}}    & x_{0I}-jx_{1Q} &-x_{1I}-jx_{0Q}
\end{array}\right],
\end{eqnarray*}

\begin{eqnarray*}
L_5=\left[\begin{array}{rrrrr}
  \frac{x_0^*}{\sqrt{2}}& -\frac{x_0^*}{\sqrt{2}}&-x_1^*&-x_3^*&-x_6^*\\
 -\frac{x_0^*}{\sqrt{2}}& -\frac{x_0^*}{\sqrt{2}}&-x_2^*&-x_4^*&-x_7^*\\
   x_{1I}-jx_{2Q}       &-x_{2I}-jx_{1Q}         &\frac{x_0}{\sqrt{2}}&-\frac{x_5^*}{\sqrt{2}}
              &-\frac{x_8^*}{\sqrt{2}}\\
   x_{2I}-jx_{1Q}       & x_{1I}+jx_{2Q}&-\frac{x_0}{\sqrt{2}}& -\frac{x_5^*}{\sqrt{2}}
              & -\frac{x_8^*}{\sqrt{2}}\\
   x_{3I}-jx_{4Q}       &-x_{4I}-jx_{3Q}& \frac{x_5^*}{\sqrt{2}}& \frac{x_0}{\sqrt{2}}
              & -\frac{x_9^*}{\sqrt{2}}\\
   x_{4I}-jx_{3Q}       &x_{3I}+jx_{4Q} & \frac{x_5^*}{\sqrt{2}}&-\frac{x_0}{\sqrt{2}}
              &-\frac{x_9^*}{\sqrt{2}}\\
   \frac{x_5}{\sqrt{2}} & \frac{x_5}{\sqrt{2}}& -x_3& x_1&    0\\
   \frac{x_5}{\sqrt{2}} &-\frac{x_5}{\sqrt{2}}& -x_4& x_2&    0\\
   x_{6I}-jx_{7Q}       &-x_{7I}-jx_{6Q}& \frac{x_8^*}{\sqrt{2}}&  \frac{x_9^*}{\sqrt{2}}
              & \frac{x_0}{\sqrt{2}}\\
   x_{7I}-jx_{6Q}       & x_{6I}+jx_{7Q}& \frac{x_8^*}{\sqrt{2}}& \frac{x_9^*}{\sqrt{2}}
              &-\frac{x_0}{\sqrt{2}}\\
   \frac{x_8}{\sqrt{2}} & \frac{x_8}{\sqrt{2}}& -x_6& 0&  x_1\\
   \frac{x_8}{\sqrt{2}} &-\frac{x_8}{\sqrt{2}}& -x_7& 0&  x_2\\
   \frac{x_9}{\sqrt{2}} & \frac{x_9}{\sqrt{2}}& 0& -x_6&  x_3\\
   \frac{x_9}{\sqrt{2}} &-\frac{x_9}{\sqrt{2}}& 0& -x_7&  x_4\\
      0                 &  0                  & x_9& -x_8& x_5
\end{array}\right].
\end{eqnarray*}

\caption{The maximal rate codes for $4$ and $5$ transmit antenna with fewer zeros}
\label{c1fig4}

\end{figure*}

\begin{thebibliography}{160}

\bibitem{TJC} V. Tarokh, H. Jafarkhani, and A. R. Calderbank,
``Space-time block codes from orthogonal designs,''
{\it{IEEE Trans. Inform. Theory,}} vol. 45, pp. 1456-1467, July 1999.

\bibitem{TiH} O.Tirkkonen and  A.Hottinen,
``Square matrix embeddable STBC for complex signal constellations  for complex signal constellations Space-time block codes from orthogonal design,''
{\it{IEEE Trans. Inform. Theory,}} Vol 48,no. 2, pp. 384-395, Feb. 2002.

\bibitem{Lia} X.B.Liang
``Orthogonal Designs with Maximal Rates,''
{\it{IEEE Trans.Inform. Theory,}} Vol.49, pp. no. 10, 2468-2503, Oct. 2003

\bibitem{KhR} Zafar Ali Khan and B. Sundar Rajan,
``Single-Symbol Maximum-Likelihood Decodable Linear STBCs,''
{\it{IEEE Transactions on Information Theory,}} Vol.52, No.5, May 2006, pp.2062-2091.                                                                                        
\bibitem{ALP} J. F. Adams, P. D. Lax, and R. S. Phillips, ``On matrices whose real linear combinations are nonsingular,''
{\it{ Proc. Amer. Math. Soc.,}} vol. 16, 1965, pp. 318-322.

\bibitem{LFX} Kejie Lu, Shengli Fu and Xiang-G Xia,
``Closed-Form Designs of Complex Orthogonal Space-Time Block Codes of Rates $\frac{k+1}{2k}$ for $2k-1$ or $2k$ Transmit Antennas,''
{\it{IEEE Trans. Inform. Theory,}} vol. 51, No.5, pp. 4340-4347, Dec 2005.

 \bibitem{GP} A. V. Geramita and N. J. Pullman, ``A theorem of  Hurwitz and Radon and Orthogonal design ,''
{\it{ Proc. Amer. Math. Soc.,}} vol. 42, No. 1, 1974, pp. 51-56.

\bibitem{ChS}
Jinhui Chen and Dirk T. M. Slock,
``Orthogonal Space-Time Block Codes for Analog Channel Feedback,'' 
Proceedings of IEEE International Symposium on Information Theory, (ISIT 2008), Toronto, Canada, July 6-11, 2008, pp.1473-1477.

\bibitem{TWMS} L. C. Tran, T. A. Wysocki, A. Mertins and J. Seberry, Complex Orthogonal Space-Time Processing in Wireless Communications, Springer-Verlag, 2006.

\bibitem{DaR1} Smarajit Das and B. Sundar Rajan,
``Square Complex Orthogonal Designs with Low PAPR,''
Proceedings of IEEE International Symposium on Information Theory, (ISIT 2007), Nice, France, June 24-29, 2007, pp. 2626-2630. 

\bibitem{DaR2} Smarajit Das and B. Sundar Rajan,
``Square Complex Orthogonal Designs with Low PAPR and Signaling Complexity,''
To appear in IEEE Transactions on Wireless Communications. Also available as arXiv:0807-4128v1 [cs.IT] 25 Jul 2008.

\end{thebibliography}
\end{document}